\documentclass{aa}
\usepackage{txfonts}
\usepackage{graphicx}

\begin{document}

\title{A JOINT MID-INFRARED SPECTROSCOPIC AND X-RAY IMAGING INVESTIGATION OF LINER GALAXIES}

\author{S. Satyapal \inst{1,2} \and R. M. Sambruna \inst{1,3} \and R. P. Dudik \inst {1}} 

\offprints{S. Satyapal, \email{satyapal@physics.gmu.edu}}

\institute{George Mason University, Department of Physics \& Astronomy, MS 3F3, 4400 University Drive, Fairfax, VA 22030 \and Presidential Career Award Scientist \and George Mason University, School of Computational Sciences, MS 5C3, 4400 University Drive, Fairfax, VA 22030 }

\date{Received Date/Accepted Date}

\abstract{We present a comprehensive comparative high resolution mid-IR spectroscopic and X-ray imaging investigation of LINERs using archival observations  from the {\it ISO}-SWS and the {\it Chandra} Advanced CCD Imaging Spectrometer.  Although the sample is heterogenous and incomplete, this is the first comprehensive study of the mid-infrared fine structure line emission of LINERs.  These results have been compared with similar observations of starburst galaxies and AGN.  We find that LINERs very clearly fall between starbursts and AGN in their mid-IR fine structure line spectra, showing L$_{[OIV]26{\mu}m}$/L$_{FIR}$ and L$_{[OIV]26{\mu}m}$/L$_{[NeII]12.8{\mu}m}$ ratios, both measures of the dominant nuclear energy source in dust-enshrouded galaxies, intermediate between those of AGN and starbursts. {\it Chandra} imaging observations of the LINERs reveal hard nuclear point sources morphologically consistent with AGN in most (67\%) of the sample, with a clear trend with IR-brightness. Most LINERs that show a single dominant hard compact X-ray core are IR-faint ((L$_{FIR}$/L$_{B}$ $<$ 1), whereas most LINERs that show scattered X-ray sources are IR-bright.  A comparative X-ray/mid-IR spectroscopic investigation of LINERs reveals some puzzling results.  Objects that display strong hard nuclear X-ray cores should also display high excitation lines in the IR.  However, we find two LINERs disagree with this expectation. The galaxy NGC 404 shows weak soft X-ray emission consistent with a starburst but has the most prominent highest excitation mid-IR spectrum of our entire sample.  Using IR emission line diagnostics alone, this galaxy would be classified as hosting a dominant AGN.  Conversely, the IR luminous LINER NGC 6240 has an extremely luminous binary AGN as revealed by the X-rays but shows weak IR emission lines.  With the advent of SIRTF, and future IR missions such as Herschel and JSWT, it is increasingly critical to determine the origin of these multiwavelength anomalies.

}
\titlerunning{IR and X-ray Study of LINERs}
\authorrunning{S. Satyapal et al.}
\maketitle

\section{INTRODUCTION AND MOTIVATION}

Of the various classes of galaxies, low-ionization nuclear emission-line region galaxies (LINERs) represent one of the most captivating subsets of active galaxies yet they are the least understood.  These galaxies show narrow optical emission lines of low ionization uncharacteristic of photoionization from normal stars (Heckman et al. 1980).  The ionization mechanism responsible for these line ratios is still controversial and remains at the forefront of astrophysics today.  LINERs constitute as much as one third of all nearby galaxies (e.g., Ho, Filippenko, \& Sargent 1997) and nearly half of all nearby early-type galaxies (e.g., Stauffer 1982).  Since Seyfert galaxies constitute only at most a few percent of the low luminosity active galactic nuclei (AGN) population, LINERs can potentially dominate the population of the low end of the AGN luminosity function and therefore can have a tremendous impact on several key critical issues ranging from the contribution of AGN to the infrared (IR) and X-ray background radiation to the origin and evolution of galaxies in general.

Many LINERs are expected to contain obscured nuclei.  In fact, using the compilation by Carillo  et al. (1999) of all LINERs studied in the literature, IR-bright LINERs (L$_{FIR}$/L$_{B}$$>$1) constitute nearly 80\% of all nearby LINERs.  In such galaxies, mid-IR spectroscopic and high spatial resolution X-ray observations are ideal probes of possible buried AGN cores. Mid-IR lines not only penetrate dust-enshrouded regions, but they also provide powerful tools to discriminate between gas photoionized by a central AGN or young stars, or shock-excited gas.  Likewise, sensitive hard X-ray observations at high spatial resolution can provide a definite probe of obscured AGN, out to column densities of a few times 10$^{24}$ cm$^{-2}$.  Detection of a single compact hard X-ray source coincident with the nucleus would constitute strong evidence for the existence of a central black hole.  Those galaxies that reveal a compact hard X-ray nuclear source should also display high excitation fine structure lines in the mid-infrared, where dust obscuration is minimal.

There have been several recent optical spectroscopic and X-ray studies of nearby optically-selected LINER galaxies (e.g., Ho et al. 1993; 1997a,b; 2001, Eracleous et al. 2002).  In these studies, detection of broad H$\alpha$ emission in some LINERs, observations of extended LINER emission in a number of galaxies, and X-ray morphology all indicate heterogeneity in the ionization mechanism responsible for the LINER spectra.  Viable explanations include shock heating through cloud-cloud collisions in galaxy mergers or surrounding AGN, cooling flows, and photoionization by nonthermal energy sources or extremely hot stars (see review by Filipenko 1996).  Because the nuclear regions of many LINERs can be obscured and the intrinsic continuum from the central energy source is often thermally reprocessed by dust and hence not directly observable,  a definitive consensus on the true origin of the observed optical emission line spectrum in all LINERs has not been reached.

\begin{table*}
\begin{center}
\begin{tabular}{llcccccc}
\multicolumn{8}{l}{{\bf Table 1: The {\it ISO}-SWS and {\it Chandra} Sample}} \\
& &\\ \hline \hline
& & & & & & & \\
\multicolumn{1}{c}{Galaxy} & \multicolumn{1}{c}{Observed} & \multicolumn{1}{c}{Distance} & Hubble & log(L$_{FIR}$) & L$_{FIR}$/L$_{B}$ & Broad & N$_H$  \\ 
\multicolumn{1}{c}{Name} & \multicolumn{1}{c}{By:} & \multicolumn{1}{c}{(Mpc)} & Type & & & H$\alpha$$?$ & $\times$10$^{22}$cm$^{-2}$ \\ 

\multicolumn{1}{c}{(1)} & \multicolumn{1}{c}{(2)} & \multicolumn{1}{c}{(3)} & (4) & (5) & (6) & (7) & (8) \\
& & & & & & & \\ \hline
& & & & & & & \\
NGC0224(M31) & {\it Chandra}, SWS, LWS & 0.8$^a$ & SA(s)b & 9.0 & 0.5 & $\cdots$ & 0.067 \\
NGC0253 & {\it Chandra}, SWS, LWS & 2.6$^b$ & SAB(s)c & 10.1 & 9.8 & $\cdots$ & 0.014 \\
NGC0404 & {\it Chandra}, SWS, LWS & 2.4$^c$ & SA(s)0 & 7.3 & 0.6 & $\cdots$ & 0.053 \\
IRAS01173+1405 &  SWS & 124.9 & $\cdots$ & 11.3 & 68.5 & $\cdots$ & 0.039  \\
NGC0660 & {\it Chandra}, SWS, LWS & 11.3$^d$ & SB(s)a;pec & 10.1 & 34.4 & $\cdots$ & 0.049 \\
3ZW 35 &  SWS, LWS  & 112.6 & $\cdots$ & 11.3 & 381.7 & $\cdots$ & 0.051 \\
NGC0838  & {\it Chandra}, SWS & 54.3 & SAB(r)ab;pec & 10.5 & $<$9.0 & $\cdots$ & 0.022 \\
NGC1052 & {\it Chandra}, SWS, LWS & 29.6$^d$  & E4 & 8.8 & 0.3 & yes & 0.031 \\
AN0248+43A & {\it Chandra}, LWS & 205.2 & Gpair & 11.5 & 138.3 & $\cdots$ & 0.102 \\
UGC05101 & {\it Chandra}, SWS & 157.6 & S$?$ & 11.7 & 118.5 & $\cdots$ & 0.027 \\
NGC3031 & {\it Chandra}, LWS & 3.6$^e$ & SA(s)ab & 8.4 & 0.1 & yes & 0.042 \\
NGC3079 & {\it Chandra}, SWS & 15.0 & SB(s)c & 10.3 & 16.6 & yes  & 0.008 \\
NGC3368(M96) & {\it Chandra}, LWS & 12.0 & SAB(rs)ab & 9.4 & 1.0 & $\cdots$ & 0.028 \\
NGC3623(M65) & {\it Chandra}, LWS & 10.8 & SAB(rs)a & 9.0 & 0.5 & $\cdots$ & 0.025 \\
NGC4125 & {\it Chandra}, LWS & 18.1 & E6;pec & 8.6 & 0.1 & $\cdots$ & 0.018 \\
NGC4278 & {\it Chandra}, LWS & 16.1$^e$ & E1-2 & 7.9 & 0.2 & yes & 0.018 \\
NGC4314 & {\it Chandra}, LWS & 12.8 & SB(rs)a & 9.0 & 1.2 & $\cdots$ & 0.018 \\
NGC4374(M84) & {\it Chandra}, LWS & 18.4$^e$ & I & 8.2 & 0.3 & $\cdots$ & 0.026 \\
NGC4486(M87) & {\it Chandra}, LWS & 16.0$^f$ & E+0-1;pec & $\cdots$ & $<$0.03 & $\cdots$ & 0.025 \\
NGC4569(M90) & {\it Chandra}, SWS, LWS & 16.8$^e$ & SAB(rs)ab & 9.7 & 1.1 & $\cdots$ & 0.025 \\
NGC4579(M58) & {\it Chandra}, SWS, LWS & 16.8$^e$ & SAB(rs)b & 9.7 & 0.9 & yes & 0.025 \\
NGC4651 &  SWS, LWS  & 10.7 & SA(rs)c & 9.1 & 2.1 & $\cdots$ & 0.020 \\
NGC4696 & {\it Chandra}, SWS, LWS & 39.5$^d$ & E+1;pec & 8.8 & 0.1 & $\cdots$ & 0.081 \\
UGC08335 &  SWS & 124.3 & Sc & 11.4 & 77.5 & $\cdots$ & 0.019 \\
UGC8387 &  SWS, LWS & 93.4 & IM;pec & 11.3 & 73.8 & $\cdots$ & 0.100 \\
NGC5194(M51) & {\it Chandra}, SWS, LWS & 8.4$^g$ & SA(s)bc;pec & 9.5 & 1.7 & yes & 0.016 \\
NGC5195 & {\it Chandra}, LWS & 7.7$^e$ & SB01;pec & $\cdots$ & $<$3.0 & $\cdots$ & 0.016 \\
MRK273 & {\it Chandra}, SWS, LWS & 151.11 & Ring galaxy & 11.8 & 162.9 & $\cdots$ & 0.011 \\
CGCG162-010 & {\it Chandra}, SWS, LWS & 253.0& cD;S0$?$ & $\cdots$ & $<$3.5 & $\cdots$ & 0.012 \\
NGC5899 &  SWS & 34.2 & SAB(rs)c & 9.9 & 3.8 & $\cdots$ & 0.017 \\
NGC6240 & {\it Chandra}, SWS, LWS & 97.9 & I0;pec & 11.5 & 38.2 & yes & 0.058 \\
NGC6503 & {\it Chandra}, LWS & 0.6 & SA(s)cd & 6.8 & 2.1 & $\cdots$ & 0.041 \\
NGC6500 & {\it Chandra}, SWS & 40.1 & SAab & 9.4 & 1.4 & $\cdots$ & 0.074 \\
NGC6764 &  SWS, LWS  & 32.2 & SB(s)bc & 10.0 & 5.8 & $\cdots$ & 0.061 \\
NGC7331 & {\it Chandra}, LWS & 11.0 & SA(s)b & 9.8 & 3.8 & $\cdots$ & 0.086 \\
NGC7479 &  SWS & 31.8 & SB(s)c & 10.3 & 4.9 & $\cdots$ & 0.050 \\
IC1459 & {\it Chandra}, LWS & 22.6 & E3 & 8.6 & 0.1 & $\cdots$ & 0.012 \\
Abell 2597 & {\it Chandra}, SWS & 328.8 & E & $\cdots$ & $\cdots$ & $\cdots$ & 0.025 \\
IRAS23128-5919 & {\it Chandra}, SWS, LWS & 178.4 & Merger & 11.7 & 89.6 & $\cdots$ & 0.028 \\
IRAS23135+2516 & SWS & 109.4 & Spiral & 11.2 & 67.8 & $\cdots$ & 0.056 \\
IRAS20551-4250 & {\it Chandra}, SWS & 171.3 & Merger & 11.7 & 66.6 & $\cdots$ & 0.039 \\
MRK266 & {\it Chandra}, SWS & 112.2 & Compact;pec & 11.2 & 25.8 & $\cdots$ & 0.017 \\
& & & & & & & \\ \hline 
\end{tabular}
\end{center}
{\scriptsize{\bf Columns Explanation:} Col(1):Common Source Names; Col(2): Sources observed by either {\it Chandra}, SWS or LWS;  Col(3):  Distance (for H$_0$= 75 km s$^{-1}$Mpc$^{-1}$ unless otherwise noted); Col(4): Morphological Class; Col(5): Far-infrared luminosities (in units of solar luminosities: L$_{\odot}$) correspond to the 40-500$\mu$m wavelength interval and were calculated using the IRAS 60 and 100 $\mu$m fluxes according to the prescription: L$_{FIR}$=1.26$\times$10$^{-14}$(2.58f$_{60}$+f$_{100}$) in W m$^{-2}$ (Sanders \& Mirabel 1996).; Col(6):  L$_{FIR}$ same as previous (Sanders \& Mirabel 1996), L$_{B}$: B magnitude see Carrillo et al (1999); Col(7): LINERs with broad H$\alpha$ emission; Col(8): N$_H$$\times$10$^{22}$cm$^{-2}$} 
{\noindent{\scriptsize{\bf References:} $^a$Stanek et al 1998; $^b$Puche et al 1988; $^c$Tully et al 1998; $^d$Guainazzi 2000; $^e$Ho et al 2001; $^f$Wilson et al 2001; $^g$Feldmeier et al 1997.}}

\end{table*}


There have been very few IR spectroscopic and high spatial resolution X-ray imaging studies of nearby LINERs.  While recent results from the Infrared Space Observatory ({\it ISO}, Kessler et al. 1996) have yielded a wealth of information on the mid-IR spectral signatures of a significant number of optically-identified standard AGN (e.g., Sturm et al. 2002) and starburst galaxies (e.g., Genzel et al. 1998, Thornely et al. 2000), very few mid-IR spectroscopic observations of LINER galaxies have yet been published.  Of the few near- and mid-IR spectroscopic observations of LINERs that have been published thus far, almost all are IR-bright and virtually all show low resolution spectra similar to starburst galaxies (e.g., Veilleux, Sanders, \& Kim 1999; Lutz et al. 1999).  Does this trend extend to all LINER galaxies? Do infrared-bright LINERs represent a "shock-dominated" subclass of the LINER population powered predominantly by star formation? As an added puzzle, these spectral characteristics are in at least one case irreconcilable with X-ray observations that reveal distinct evidence for a hard radiation field characteristic of an AGN . The optical spectrum of the strongly interacting system NGC 6240 shows LINER-like emission lines and the mid-IR properties show only weak high excitation line emission and an overall spectrum similar to starburst galaxies (Lutz, Veilleux, \& Genzel 1999). However, a number of hard X-ray observations of NGC 6240  with {\it XMM}, {\it ROSAT}, {\it Beppo SAX}, and {\it Chandra} have detected the presence of a heavily obscured AGN with total intrinsic luminosity in the QSO range (Komossa et al. 2003; Keil, Boller, \& Fujimoto 2001; Komossa \& Schulz 1999; Vignati et al. 1999, Komossa et al. 2002).  If this LINER is indeed powered by an AGN, why does it not display strong high excitation lines at mid-IR wavelengths where dust obscuration is minimal?  Other IR-bright LINERs can also host obscured AGN that have thus far been undiscovered in IR surveys of local and future high redshift surveys. In fact, some of the luminous hard X-ray sources recently discovered by deep {\it Chandra} surveys have optical and near-IR counterparts that show little or no evidence for AGN activity (eg., Brandt et al. 2001, Barger et al. 2001, Comastri et al. 2002).  Many of these objects could be similar to the local LINERs studied here. With the advent of {\it SIRTF}, it is critical to determine the origin of these possible multiwavelength inconsistencies.\\

In this paper, we present a comprehensive comparative high resolution mid-IR spectroscopic and X-ray imaging investigation of all LINERs observed by both {\it ISO} and {\it Chandra}.  The {\it Chandra} observations presented here are largely unpublished  archival observations of all LINERs observed either by the short wavelength spectrometer (SWS) or the long wavelength spectrometer (LWS) on board {\it ISO}.  In this paper we present only the SWS observations of all LINERs in the {\it ISO} archive, most of which have not previously been published.  In a companion paper, our archival  far-IR spectroscopic observations of all LINERs are presented (Satyapal \& Sanei 2003).  This paper is structured as follows.  In section 2, we summarize the SWS and {\it Chandra} LINER samples.  In section 3, we present the observations and data reduction together with a tabulation of all SWS line fluxes, X-ray luminosities, and X-ray morphological class designations.  The discussion of our results  including a  mid-IR spectroscopic and X-ray comparison of LINERs with standard AGN and starbursts is given in section 4,  followed by a summary of our main results and conclusion in section 5.


\begin{figure*}[]
{\includegraphics[width=8cm]{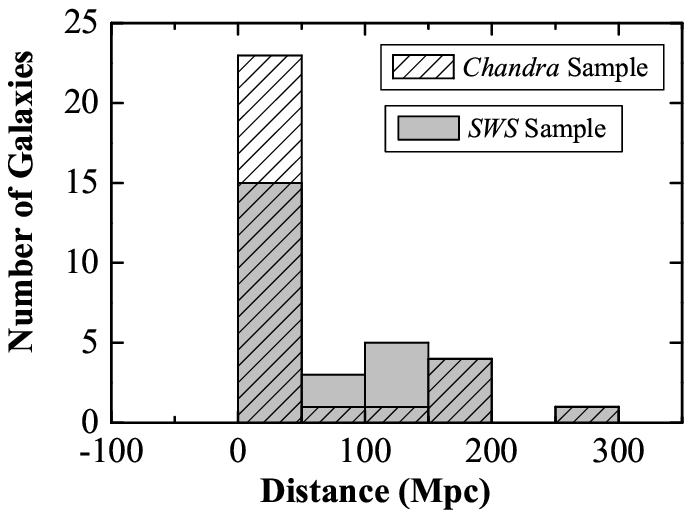}}{\includegraphics[width=8cm]{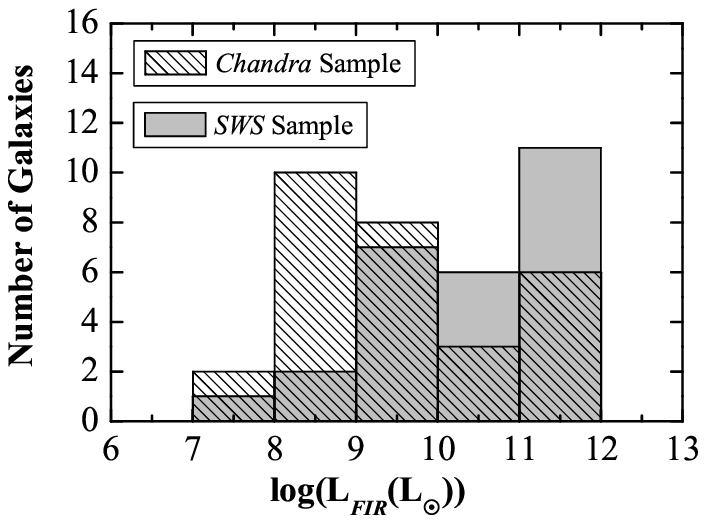}}\\
{\includegraphics[width=8cm]{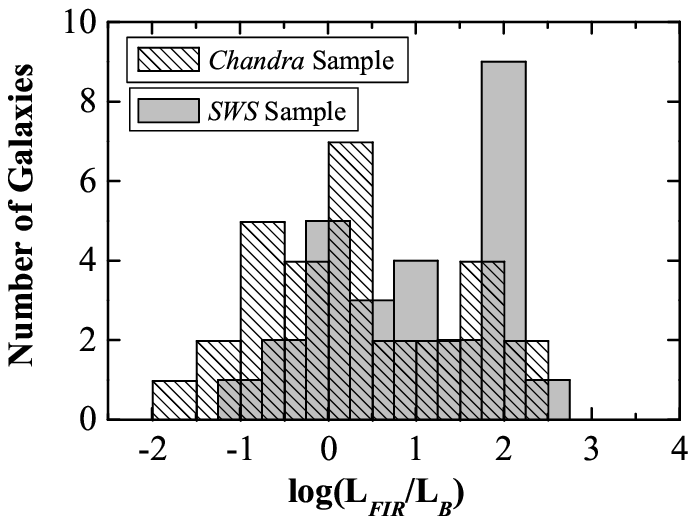}}{\includegraphics[width=8.25cm]{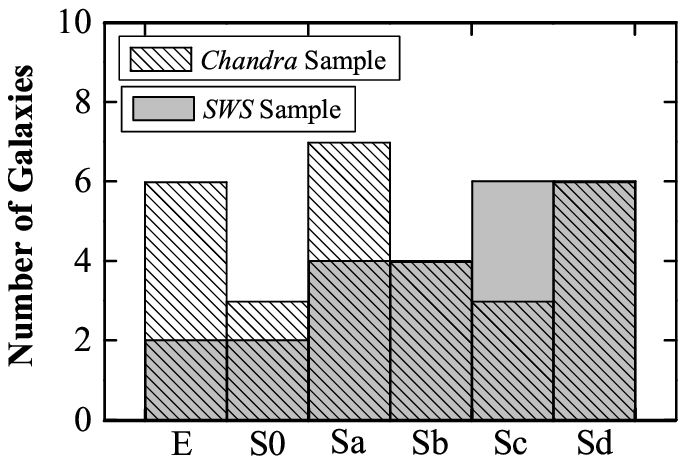}}\\
\caption[]{Characteristics of the SWS and {\it Chandra} sample of LINERs.  Most galaxies are nearby and span a wide range of luminosities, IR-brightness ratios, and Hubble types.}
\end{figure*} 

\begin{figure*}[]
\begin{center}
{\includegraphics[width=4.5cm,angle=90]{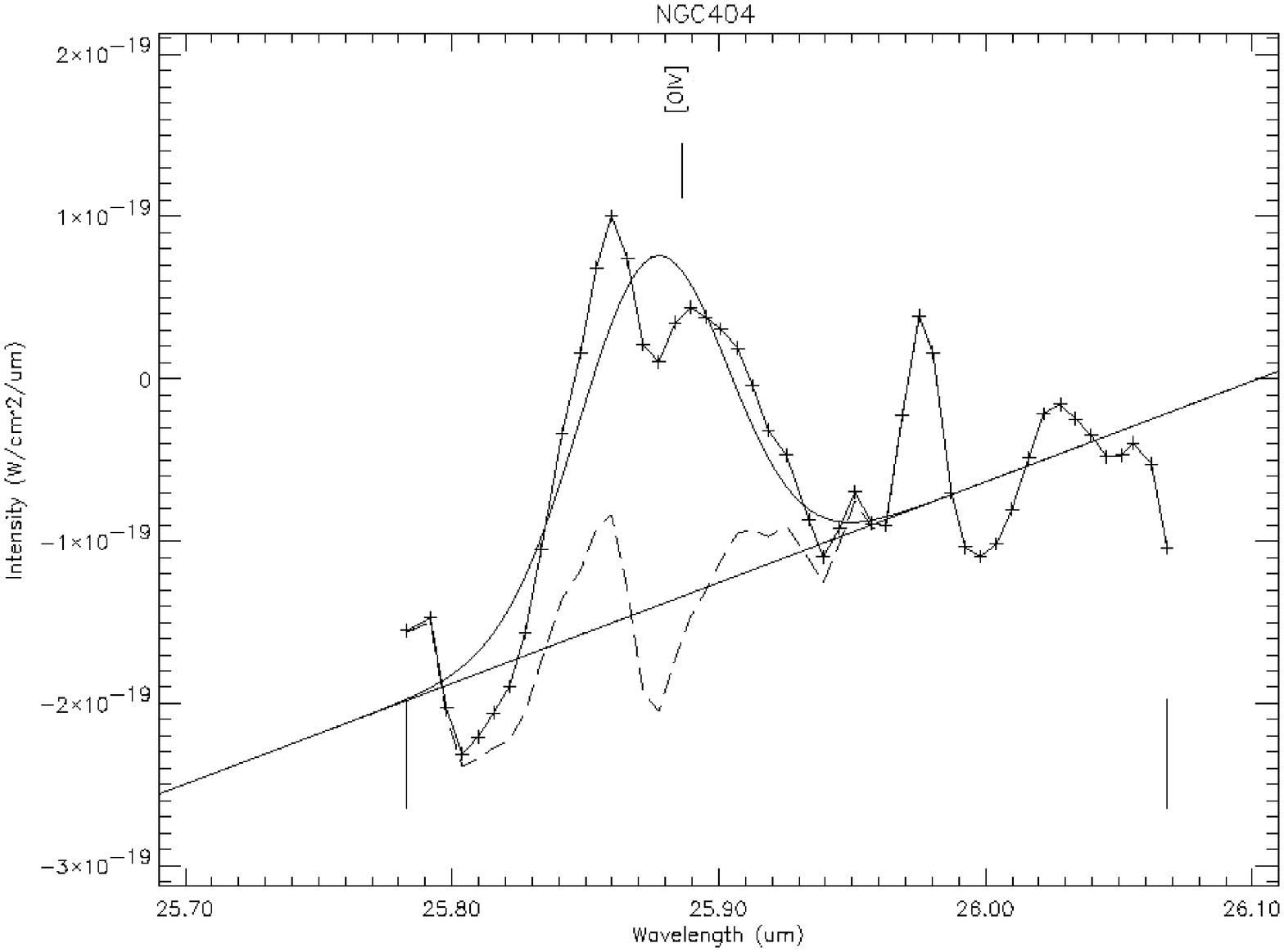}}{\includegraphics[width=4.5cm,angle=90]{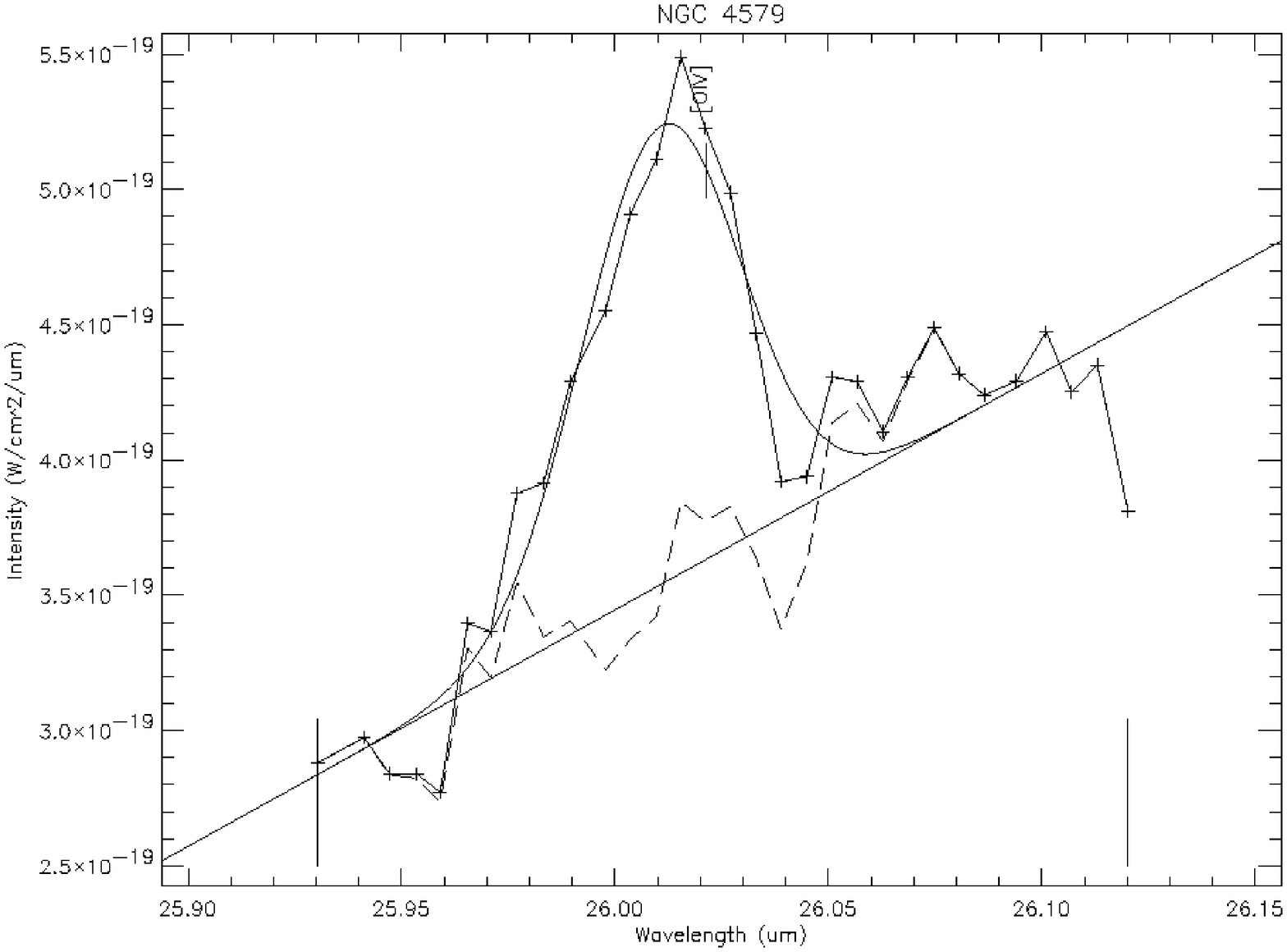}}\\

\noindent{\includegraphics[width=4.5cm,angle=90]{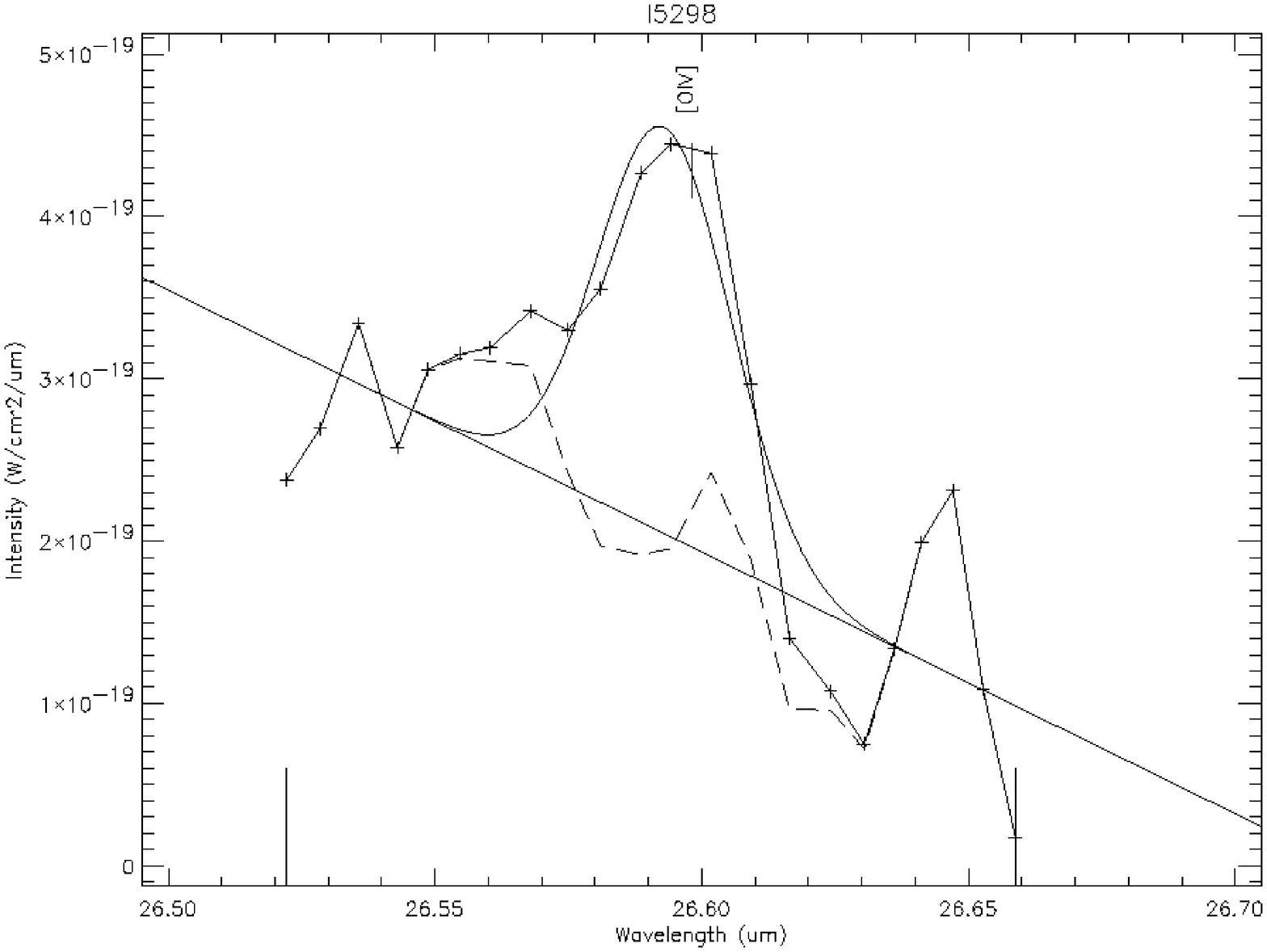}}{\includegraphics[width=4.5cm,angle=90]{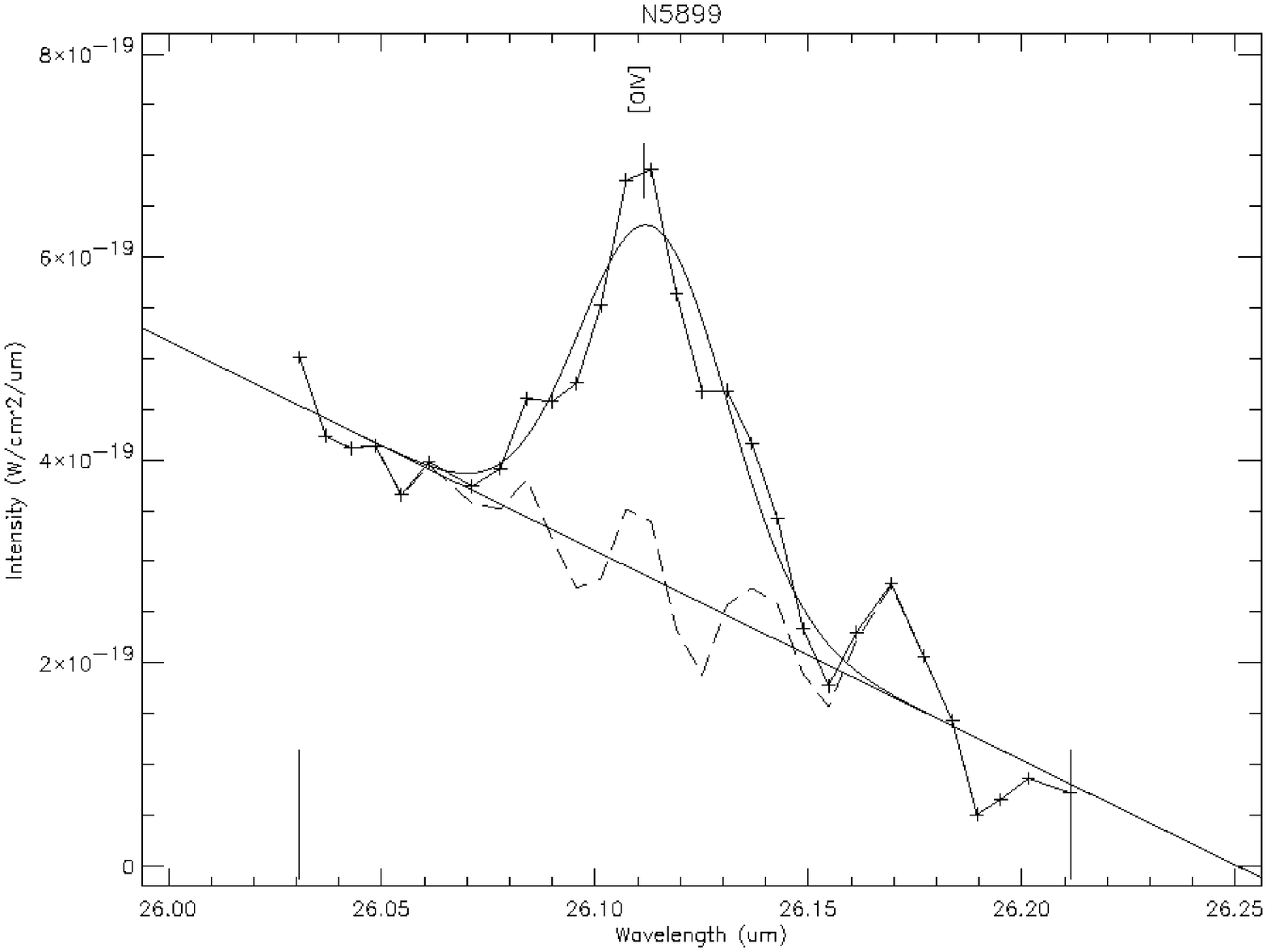}}\\

\noindent{\includegraphics[width=4.5cm,angle=90]{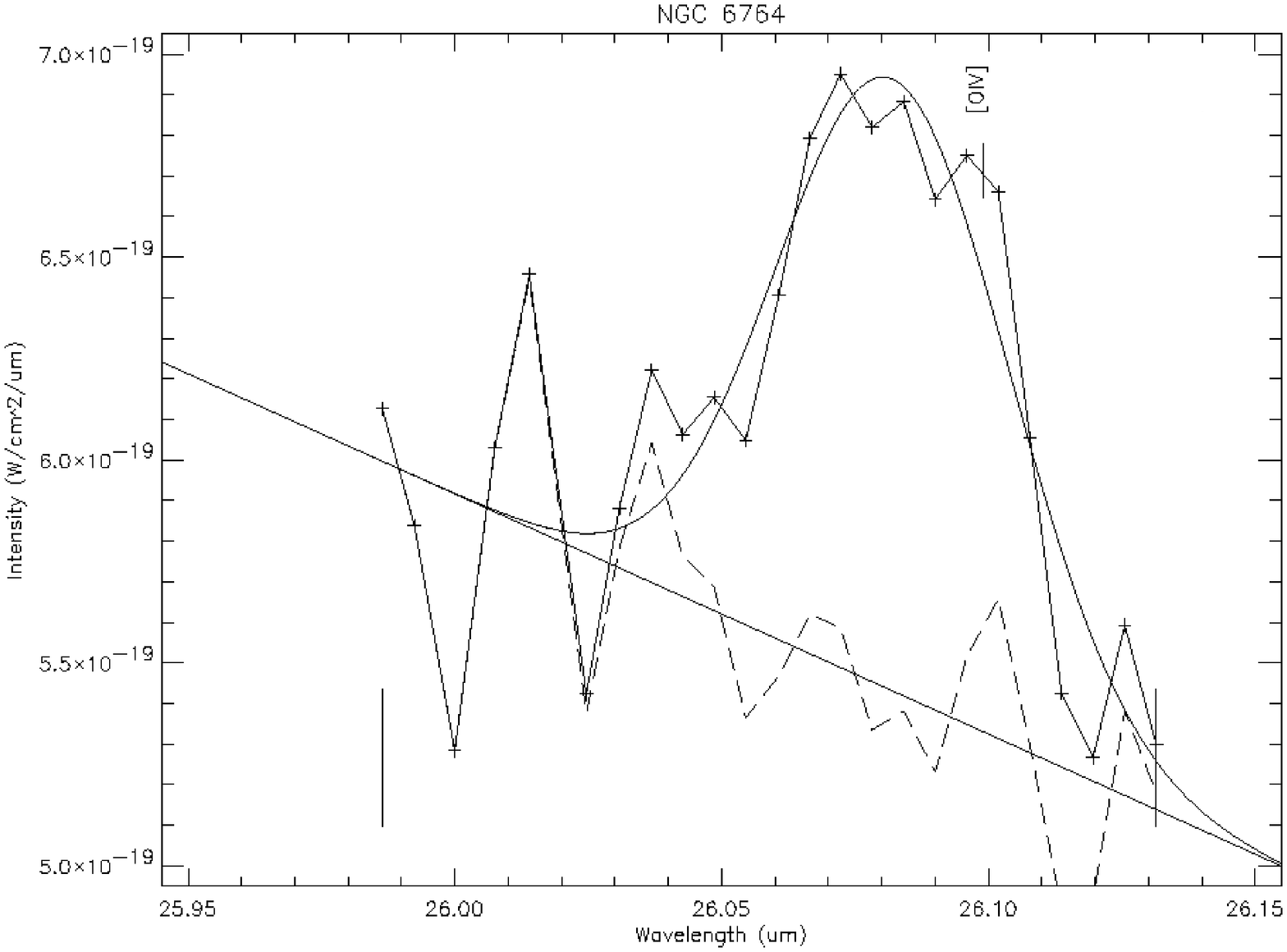}}{\includegraphics[width=4.5cm,angle=90]{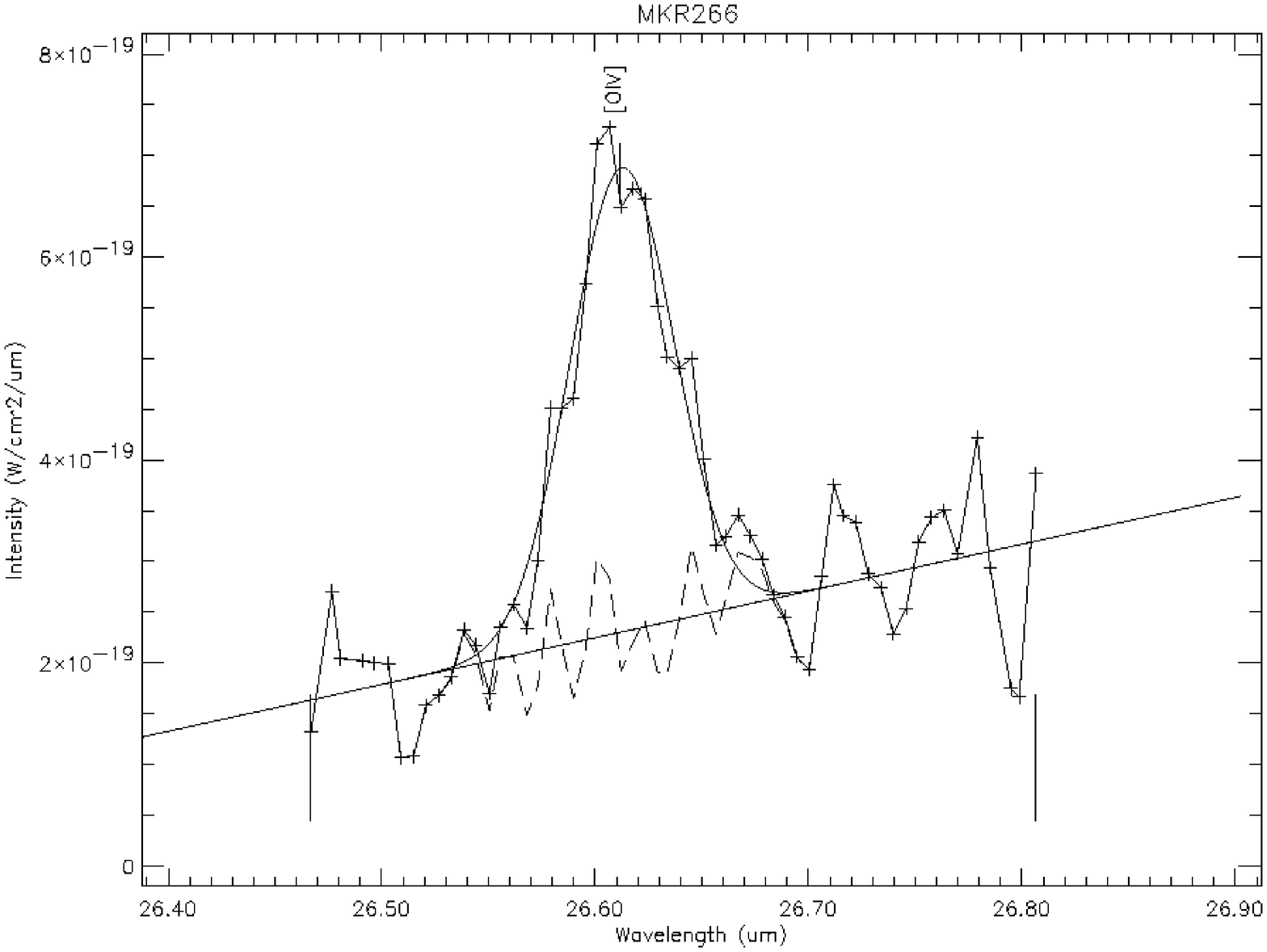}}\\

\noindent{\includegraphics[width=4.5cm,angle=90]{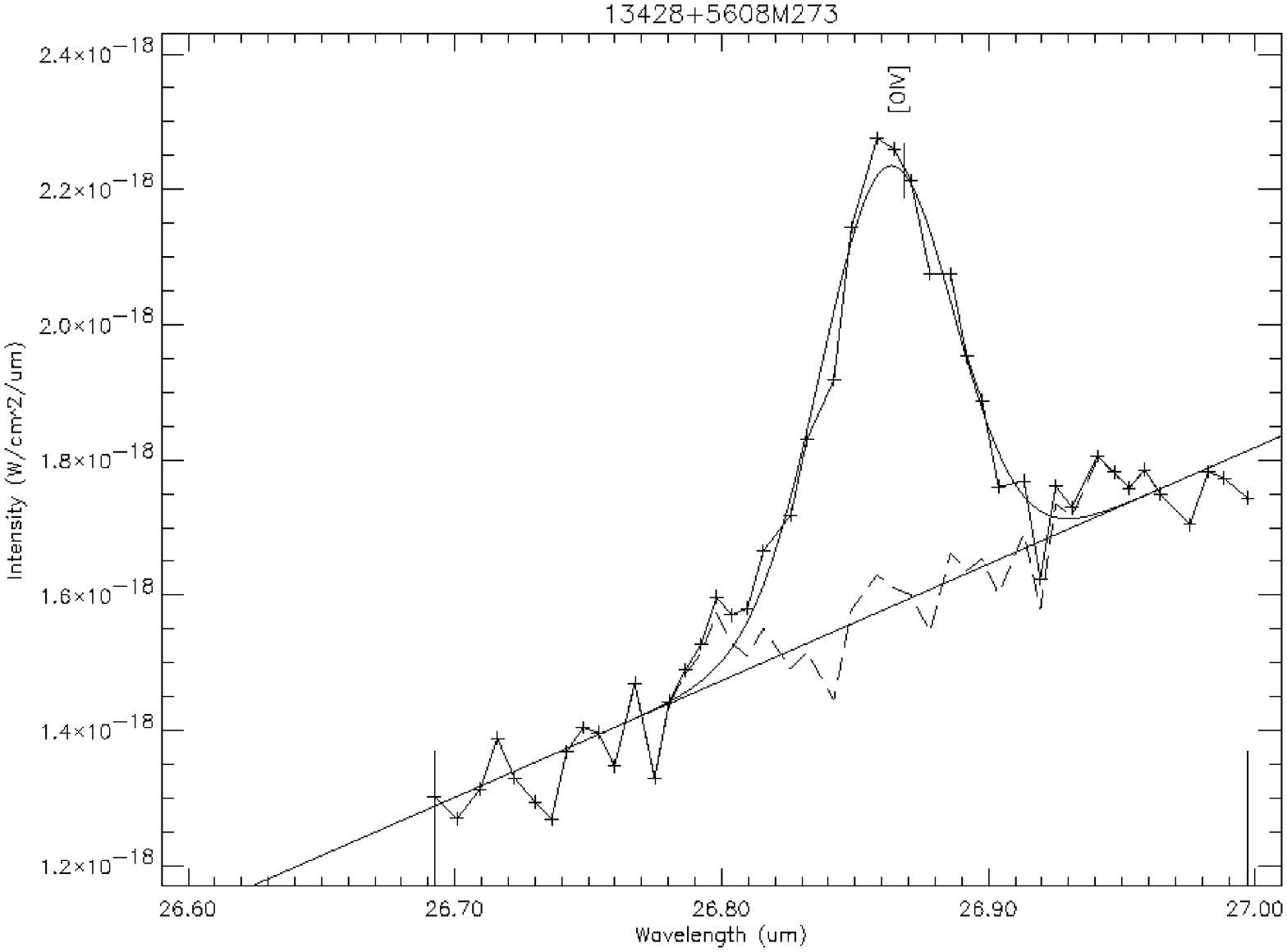}}{\includegraphics[width=4.5cm,angle=90]{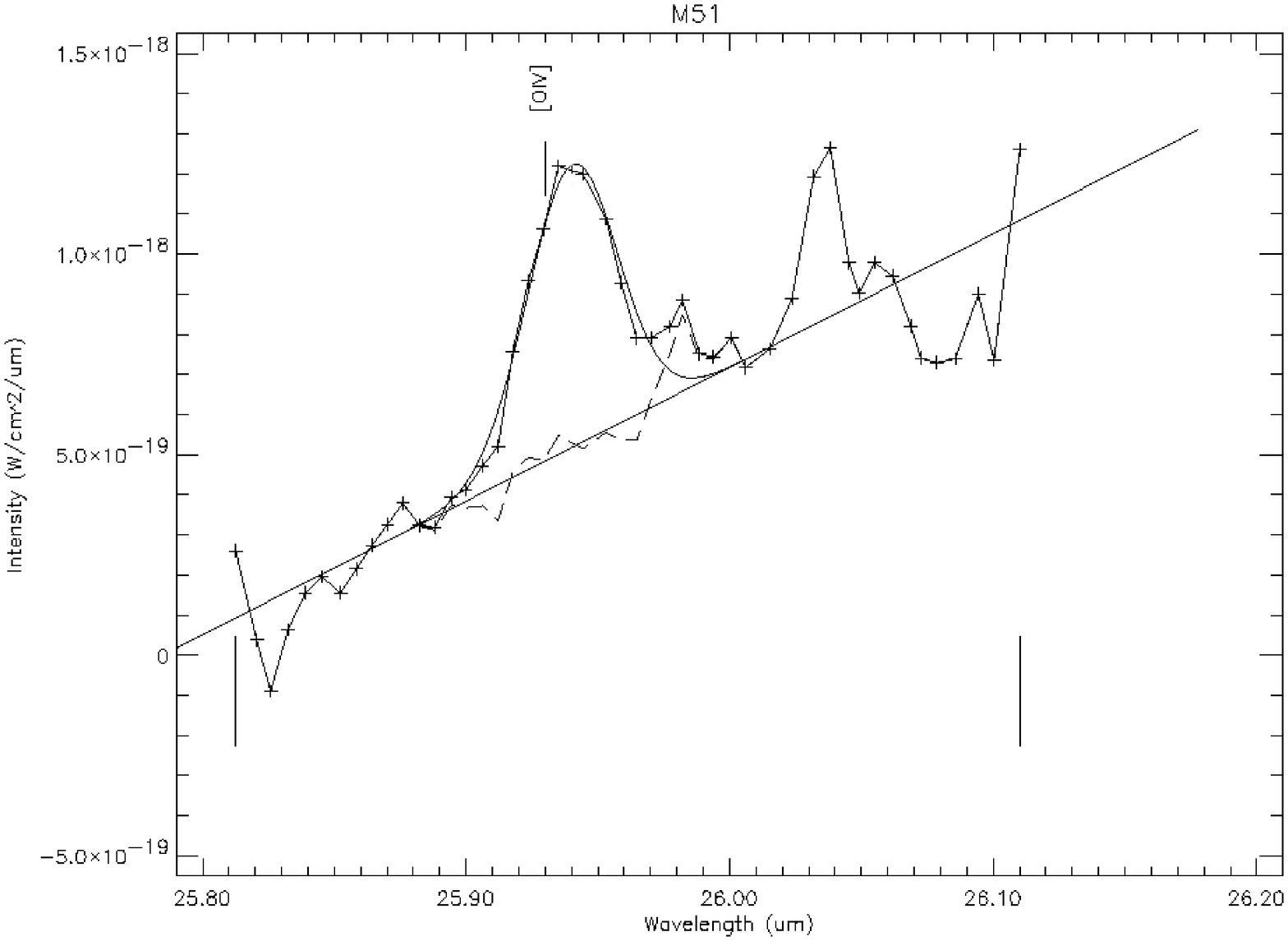}}\\

\noindent{\includegraphics[width=4.5cm,angle=90]{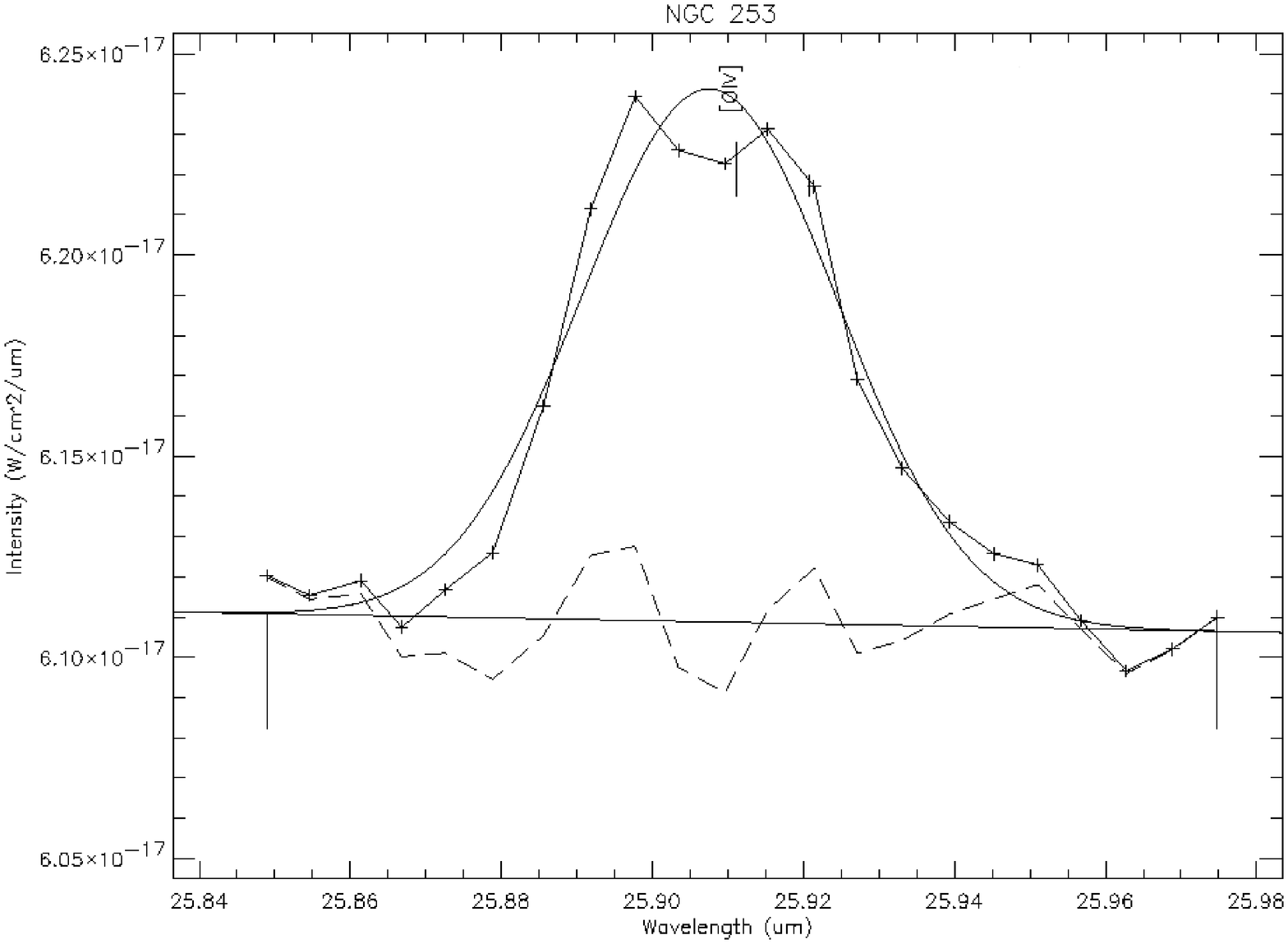}}{\includegraphics[width=4.5cm,angle=90]{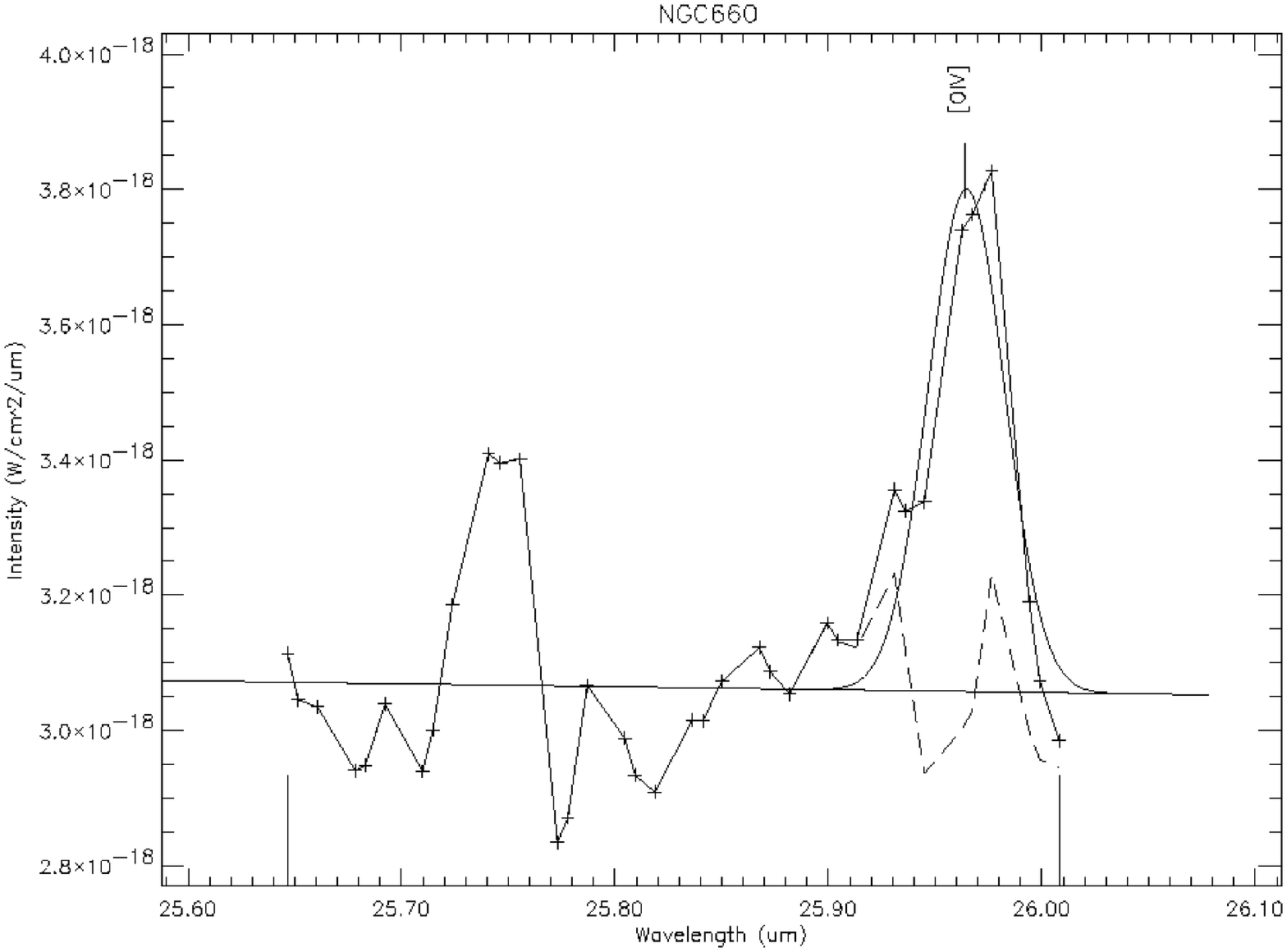}}\\
\end{center}

\caption[]{Spectra of the [OIV] 25.6 $\mu$m line, detected in 11 out of 18 LINERs observed at this wavelength for all previously unpublished data.}
\end{figure*}

\section{MID-IR AND X-RAY LINER SAMPLES}
\subsection{\it SWS LINER Sample}
We searched the {\it ISO} SWS archive for all galaxies classified as LINER galaxies using the recent compilation of LINERs by Carillo et al. (1999).  SWS observations of 28 
LINER galaxies were found to exist in the {\it ISO} archive.  Since most galaxies were observed as part of a random collection of guaranteed- and open-time programs with differing scientific goals, the sample is not complete in any sense.  Furthermore the observed wavelength range is not uniform for all galaxies. The galaxies range in distance from 0.8 to 253 Mpc, and span a wide range of infrared luminosities (L$_{FIR}$ = 6.28 $\times$ 10$^8$ L$_{\odot}$ to 6.53 $\times$ 10$^{11}$ L$_{\odot}$) , Hubble types, and infrared-to-blue luminosity ratios.  In Figure 1, we summarize the basic properties of the SWS LINER sample.  The individual galaxies observed are listed in Table 1. Basic galaxy properties were taken from  Carillo et al. which were predominantly based on the NASA/IPAC Extragalactic Database (NED).  Distances adopted were based on  Hubble constant of H$_0$ = 75 km s$^{-1}$ Mpc$^{-1}$ unless otherwise noted.


\begin{table*}
~~~~ \\
\begin{center}
\begin{tabular}{lccccc}
\multicolumn{6}{l}{{\bf Table 2a:} Observed fine structure line fluxes and upper limits (in 10$^{-20}$ W cm$^{-2}$)} \\

\hline 
\hline

\multicolumn{1}{l}{Galaxy Name} & \multicolumn{1}{c}{[SIV]} & \multicolumn{1}{c}{[NeII]} & [NeV] & [NeIII] & [SIII] \\                    
\hline
& & & & & \\
\multicolumn{1}{l}{Wavelength:} & \multicolumn{1}{c}{10.5$\mu$m} & \multicolumn{1}{c}{12.8$\mu$m} & 14.3$\mu$m & 15.5$\mu$m & 18.7$\mu$m  \\
\multicolumn{1}{l}{E$_{ion}$(eV)$^1$:} & \multicolumn{1}{c}{34.8} & \multicolumn{1}{c}{21.6} & 97.1 & 41.0 & 23.3  \\
\multicolumn{1}{l}{Aperture$^2$:} & \multicolumn{1}{c}{14$\times$20} & \multicolumn{1}{c}{14$\times$27} & 14$\times$27 & 14$\times$27 & 14$\times$27 \\
& & & & & \\
\hline
\hline
& & & & & \\

NGC0253 & $<$3.27$^3$ & 400$^a$ & 4.51$\pm$1.17 & 32.24$\pm$1.43 & $\cdots$  \\

NGC0404 & $\cdots$ & $<$0.46$^3$ & $\cdots$ & $<$2.27$^3$ & $\cdots$ \\

IRAS01173+1405 & $\cdots$ & $<$2.09$^3$ & $\cdots$ & $<$0.96$^3$ & $\cdots$ \\

NGC0660 & $\cdots$ & 30.7$\pm$2.91 & $\cdots$ & $\cdots$ & $<$7.50$^3$ \\  

NGC0838 & $\cdots$ & $\cdots$ & $\cdots$ & $\cdots$ & $\cdots$ \\

NGC1052 & $\cdots$ & 1.2$\pm$.001$^b$ & $\cdots$ & $<$0.4$^b$ & $<$0.9$^b$ \\

UGC05101 & $\cdots$ & $\cdots$ & $<$1.5$^c$ & $\cdots$ & $<$1.4$^c$\\

NGC3079 & $\cdots$ & $\cdots$ & $\cdots$ & $\cdots$ & 6.85$\pm$0.94 \\ 

NGC4579/M58 & $\cdots$ & 3.18$\pm$1.08 & $\cdots$ & 6.6$\pm$1.77 & $<$0.78$^3$ \\

NGC4651 & $\cdots$ & $\cdots$ & $<$0.45 & $\cdots$ & $\cdots$ \\

UGC08335 & $\cdots$ & $\cdots$ & $<$0.53$^3$ & $\cdots$ & $\cdots$ \\

IC0883 & $\cdots$ & $\cdots$ & $\cdots$ & $\cdots$ & $\cdots$ \\

M51 & 1.52$\pm$0.34 & 6.63$\pm$0.49 & $<$0.46$^3$ & 3.05$\pm$0.39 & 1.24$\pm$0.23 \\

MRK0266 & $\cdots$ & $\cdots$ & 0.5$^d$ & $\cdots$ & $\cdots$ \\

MRK273 & $\cdots$ & 3.0$^c$ & 0.82$^c$ & $<$1.52$^3$ & $<$0.82$^c$ \\

CGCG162-010 & $\cdots$ & $\cdots$ & $\cdots$ & $\cdots$ & $\cdots$ \\

NGC5899 & $\cdots$ & $\cdots$ & $\cdots$ & $\cdots$ & $\cdots$ \\

NGC6240 & $\cdots$ & 17$^a$ & 1.0$^c$ & 6.7$^a$ & $<$4.0$^c$ \\

NGC6500 & $\cdots$ & $<$0.57$^3$ & $\cdots$ & $<$10.9$^3$ & $<$2.52$^3$ \\

NGC6764 & $<$.52$^3$ & 7.87$\pm$.63 & 1.71$\pm$.46 & 8.61$\pm$.81 & 17.65$\pm$3.84 \\

IRAS 20551-4250 & $\cdots$ & 1.3$^c$ & $<$0.25$^c$ & $\cdots$ & 0.3$^c$ \\

NGC7479 & $\cdots$ & $\cdots$ & $\cdots$ & $\cdots$ & $\cdots$ \\

IRAS 23128-5919 & $\cdots$ & 5.65$\pm$0.30$^c$ & $<$0.4$^c$ & $\cdots$ & 0.89$^c$ \\

IC5298 & $\cdots$ & $\cdots$ & $\cdots$ & $\cdots$ & $\cdots$ \\

& & & & & \\ \hline 
\end{tabular}
\end{center}
{\scriptsize{\bf Footnotes to Table 2a:} $^1$lower ionization potential of the stage leading to the transition; $^2$in arcseconds; $^3$Upper limits correspond to 3$\sigma$ values.}

{\scriptsize{\bf References:} Flux data for these objects were taken from the following papers:   $^a$Thornley et al. 2000; $^b$ Sugai and Malkan 2000; $^c$ Genzel et al. 1998; $^d$ Prieto and Viegas 2000}
\end{table*}


\begin{table*}
~~~~ \\
\begin{center}
\begin{tabular}{lccccc}
\multicolumn{6}{l}{{\bf Table 2b: Fluxes continued} (in 10$^{-20}$ W cm$^{-2}$)} \\
 \hline \hline
\multicolumn{1}{l}{Galaxy Name} & \multicolumn{1}{c}{[OIV]} & \multicolumn{1}{c}{[FeII]} & [SIII] & [SiII] & [NeIII] \\                    
\hline
& & & & & \\
\multicolumn{1}{l}{Wavelength :} & \multicolumn{1}{c}{25.9$\mu$m} & \multicolumn{1}{c}{26.0$\mu$m}& 33.5$\mu$m & 34.8$\mu$m & 36.0$\mu$m \\
\multicolumn{1}{l}{E$_{ion}$(eV)$OIVlines/266mrk-OIV.ps^1$:} & \multicolumn{1}{c}{54.9} & \multicolumn{1}{c}{7.9} & 23.3 & 8.2 & 41.0  \\
\multicolumn{1}{l}{Aperture$^2$:} & \multicolumn{1}{c}{14$\times$20} & \multicolumn{1}{c}{14$\times$27} & 20$\times$33 & 20$\times$33 & 20$\times$33 \\
& & & & &  \\
\hline
\hline
& & & & & \\

NGC0253 & 5.63$\pm$0.46 & 19.3$\pm$1.1 & $\cdots$ & 294$\pm$7.27 & 25.3$\pm$2.64 \\

NGC0404 & 1.51$\pm$0.22 & $\cdots$ & 3.67$\pm$0.73 & 2.23$\pm$0.29 & 8.96$\pm$0.55 \\

IRAS 01173+1405 & $\cdots$ & $\cdots$ & $\cdots$ & $\cdots$ & $\cdots$ \\

NGC0660 & 3.30$\pm$0.89 & $<$2.43$^3$ & 40$\pm$3.23 & 52.8$\pm$7.2 & $\cdots$  \\  

NGC0838 & $<$0.85$^3$ & 1.63$\pm$0.28 & $\cdots$ & $\cdots$ & $\cdots$ \\

NGC1052 & $\cdots$ & $\cdots$ & $<$1.0$^{b}$ & $\cdots$ & $\cdots$ \\

UGC05101 & $<$0.600$^c$ & $\cdots$ & 2.5$^c$ & 6.6$^c$ & $\cdots$ \\

NGC3079 & $<$0.27$^3$ & & 6.62$\pm$1.74 & 32.02$\pm$3.22 & \\ 

NGC4579/M58 & 0.79$\pm$0.10 & & $<$1.19$^3$ & 13.4$\pm$1.67 & $\pm$1.17 \\

NGC4651 & $\cdots$ & $\cdots$ & $\cdots$ & $\cdots$ & $\cdots$ \\

UGC08335 & $\cdots$ & $\cdots$ & $\cdots$ & $\cdots$ & $\cdots$ \\

IC0883 & $<$0.90$^3$ & $\cdots$ & $\cdots$ & $\cdots$ & $\cdots$ \\

M51 & 1.48$\pm$0.33 & $\cdots$ & 12.1$\pm$2.71 & 12.3$\pm$2.17 & $\cdots$ \\

MRK0266 & 2.1$^d$ & $\cdots$ & $\cdots$ & $\cdots$ & $\cdots$ \\

MRK0273 & 2.8$^c$ & $\cdots$ & 2.3$^c$ & 3$^c$ & $\cdots$ \\

CGCG162-010 & $\cdots$ & $\cdots$ & $\cdots$ & $<$1.17 & $\cdots$ \\

NGC5899 & 1.5$\pm$0.17 & $\cdots$ & $\cdots$ & $\cdots$ & $\cdots$ \\

NGC6240 & 3.1$^c$ & $\cdots$ & 4.5$^c$ & 21$^c$ & $\cdots$ \\

NGC6500 & & & 2.89$\pm$0.37 & $<$0.65$^3$ & $<$2.36$^3$ \\

NGC6764 & O.83$\pm$0.19 & $<$0.31$^3$ & $<$3.28$^3$ & 13.77$\pm$1.60 & \\

IRAS 20551-4250 & $<$0.300$^c$ & $<$1.24$^3$ & 1.4$^c$ & 1.7$^c$ & $<$1.49$^3$ \\

NGC7479 & $<$0.93$^3$ & $<$0.93$^3$ & $\cdots$ & $\cdots$ & $\cdots$ \\

IRAS 23128-5919 & 0.3$^c$ & $\cdots$ & 2.8$^c$ & 3.7$^c$ & $\cdots$ \\

IC5298 & .79$\pm$.20 & $\cdots$ & $\cdots$ & $\cdots$ & $\cdots$ \\

& & & & &  \\ \hline 
\end{tabular}
\end{center}
{\scriptsize{\bf Footnotes to Table 2b:} $^1$ lower ionization potential of the stage leading to the transition; $^2$ in arcseconds; $^3$Upper limits correspond to 3$\sigma$ values.}
 
{\scriptsize{\bf References:} Flux data for these objects were taken from the following papers:  $^b$  Sugai and Malkan 2000; $^c$ Genzel et al. 1998;  $^d$ Prieto and Viegas 2000}
\end{table*}

\begin{table*}

\begin{center}
\begin{tabular}{cccccccc}
\multicolumn{8}{l}{{\bf Table 3: X-Ray Observational Details}} \\
& &\\ \hline \hline
& & & & & & & \\
\multicolumn{1}{c}{Galaxy Name} & \multicolumn{1}{c}{X-ray Class} & \multicolumn{1}{c}{L$_{X}$} & Start  & Exposure & Hard Counts & Count Rate & Counts/Frame \\ 
\multicolumn{1}{c}{} & \multicolumn {1}{c}{} & \multicolumn{1}{c}{in ergs s$^{-1}$} & \multicolumn{1}{c}{Date} & \multicolumn{1}{c}{Time} & \multicolumn{1}{c}{in Nucleus} &  0.2-10keV & \\ 

\multicolumn{1}{c}{} & \multicolumn{1}{c}{} & \multicolumn{1}{c}{} & \multicolumn {1}{c}{} & \multicolumn{1}{c}{(ks)} & \multicolumn{1}{c}{2-10keV} &  counts/sec & Pileup if $>$0.1 \\ 
\multicolumn{1}{c}{(1)} & \multicolumn{1}{c}{(2)} & \multicolumn{1}{c}{(3)} & (4) & (5) & (6) & (7) & (8) \\
& & & & & & & \\ \hline

NGC0224 & II & 3.9E+37$^a$ & 6/1/2000 & 5.16 & 51 & 0.1242 & 0.4023$^*$ \\

NGC0253 & II & 1.1E+38$^a$ & 12/16/1999 & 14.17 & 254 & 0.04161 & 0.1348$^*$ \\

NGC0404	& I & 2.1E+37$^a$ & 12/19/1999 & 24.17 & 25 & 0.0076 & 0.0246 \\
NGC0660	& II & $<$9.5e+37$^a$ & 1/28/2001 & 1.94 & $\cdots$ & $\cdots$ & $<$0.0050 \\ 
NGC0835	& II & 7.0E+39$^a$ & 11/16/2000 & 12.73 & 15 & 0.0045 & 0.0147 \\
NGC1052	& I & 4.2E+41$^b$ & 8/29/2000 & 2.4 & 183 & 0.1255 & 0.2033$^*$ \\
AN0248+43A & II & 1.8E+40$^a$ & 8/27/2000 & 31.26 & 1 & 0.0007 & 0.0003 \\ 
UGC05101 & I & 7.7E+40$^a$ & 5/28/2001 & 49.93 & 150 & 0.0072 & 0.0232 \\ 
NGC3031 & I & 1.6E+40$^c$ & 3/21/2000 & 2.41 & 76 &  0.0933 & 0.3023$^*$ \\
NGC3079 & II & 6.8E+38$^a$ & 3/7/2001 & 26.92 & 95 & 0.0078 & 0.0251 \\
NGC3368 & II & 2.8E+39$^a$ & 11/20/2000 & 2.01  & 1 & 0.0045 & 0.0147 \\
NGC3623 & IV & 4.0E+38$^a$ & 11/3/2000 & 1.76 & 2 & 0.0081 & 0.0262 \\
NGC4125 & III & 4.2E+38$^a$ & 9/9/2001 & 65.08 & 35 & 0.0042 & 0.0137 \\
NGC4278 & I & 1.2E+40$^c$ & 4/20/2000 & 1.43 & 47 &  0.2085 & 0.3378$^*$ \\
NGC4314 & II & 1.4E+38$^a$ & 4/2/2001 & 16.28 & 5 & 0.0021 & 0.0067 \\
NGC4374 & III & 1.3E+39$^c$ & 5/19/2000  & 28.85 & 153 & 0.0360 & 0.0584 \\
NGC4486 & I & 3.3E+40$^a$ & 7/29/2000 & 38.16 & 1948 & 0.3012 & 0.9759$^*$ \\
NGC4569 & II & 2.6E+39$^c$ & 2/17/2000 & 1.71 & 10 & 0.0296 & 0.0958 \\
NGC4579 & I & 8.9E+40$^c$ & 5/2/2000 & 35.64 & 8278 & 0.8062 & 0.6530$^*$ \\
NGC4696 & III & 1.3E+40$^a$ & 4/18/2001 & 85.84 & 39 & 0.0022 & 0.0072 \\
NGC5194 & III & 1.1E+41$^d$ & 6/23/2001 & 27.2 & 49 & 0.0178 & 0.0576 \\
NGC5195 & IV & 7.1E+37$^c$ & 1/23/2000 & 1.15 & $\cdots$ & 0.0009 & 0.0085 \\
MRK273 & III & 1.1E+44$^e$ & 4/19/2000 & 44.4 & 671 & 0.0264 & 0.0854 \\
CGCG162-010 & III & 7.1E+41$^a$ & 3/21/2000 & 19.67 & 52 & 0.0274 & 0.0888 \\
NGC6240 & III & 1.6E+44$^f$ & 7/29/2001 & 37.16 & 1110 & 0.0579 & 0.1878$^*$ \\
NGC6503 & II & 4.6E+35$^a$ & 3/23/2000 & 13.15 & 11 & 0.0029 & 0.0093 \\
NGC6500 & I & 1.7E+40$^a$ & 8/1/2000 & 2.13 & 1 & 0.0204 & 0.0662 \\
NGC7331 & II & 3.3E+38$^a$ & 1/27/2001 & 30.13 & 33 & 0.0051 & 0.0083 \\
IC1459 & I & 2.6E+40$^a$ & 8/12/2001 & 60.17 & 1906 & 0.1287 & 0.2084$^*$ \\
Abell 2597 & III & 1.5E+42$^a$ & 7/28/2000 & 39.86 & 142 & 0.0322 & 0.1044$^*$ \\
IRAS23128-5919 & III & 1.1E+41$^a$ & 9/30/2001 & 49.95 & 286 & 0.0109 & 0.0353 \\
IRAS20551-4250 & III & 7.2E+40$^a$ & 10/31/2001 & 45.44 & 65 & 0.0072 & 0.0232 \\
MRK266NE & I & 7.4E+40$^a$ & 11/2/2001 & 19.95 & 148 & 0.0194 & 0.0628 \\
& & & & & & & \\ 
\hline 
\end{tabular}
\end{center}
{\scriptsize{\bf Columns Explanation:} Col(1): Common Source Names ; Col(2): X-ray morphological class; Col(3): Intrinsic X-ray luminosity in the 2-10 keV band.; Col(4): Start Date of Observation; Col(5):  Actual Exposure Time in ks; Col(6): Hard Counts in 2 arcsec aperture centered on the Nucleus (2-10keV);  Col(7): Count Rate in counts per second (0.2-10keV); Col(8): Counts per Frame, $^*$ indicates pileup, pileup if counts/frame$>$0.1 }

{\noindent{\scriptsize{\bf References:} $^a$ This work.  Luminosity using PIMMS assuming an intrinsic power law slope of 1.8.  Luminosities corrected for Galacitc absorption listed in Table 1.; $^b$ Guainazzi et al. 2000; $^c$ Ho et al. 2001; $^d$ Fukazawa et al. 2001; $^e$ Xia et al. 2001; $^f$ Vignati et al. 1999.}}

\end{table*}

\subsection{\it Chandra LINER Sample}

Our sample includes all the LINERs observed by {\it Chandra} that were also observed by either the LWS or SWS for which data were available in the public archive up to 2002 November. Again, LINER identifications were based on the LINER catalog compiled by Carillo et al. (1999).  Thirty three LINER galaxies observed by the {\it ISO} spectrometers exist in the {\it Chandra} archive.  Of these galaxies, 20 were observed by the SWS and the remaining 13 were observed only by the LWS. The galaxies are mostly nearby, and span a wide range of infrared luminosities, Hubble types, and infrared-to-blue luminosity ratios.  In Figure 1, we summarize the main characteristics of our X-ray LINER sample.  The individual galaxies observed are included in Table 1.  The {\it Chandra} observations for several of the galaxies have been published before, including 6 LINERs from the Ho et al. (2001) sample (see Table 3 for references) .  In these cases, published values for X-ray luminosities and X-ray morphological class designations, if available, were adopted.  X-ray observations of 18 of the LINERs have never been published as of 2003 January, and are presented here for the first time.   


\section{OBSERVATIONS, DATA ANALYSIS, AND RESULTS}

\subsection{\it SWS Observations and Results}

All observations were obtained with the {\it ISO}-SWS (de Graauw et al. 1996) using the SWS02 or SWS06 observing modes.  The aperture size of the SWS ranges from 14\,'' $\times$ 20\,'' to  20\,'' $\times$ 33\,'' and the spectral resolving power, $\lambda$ / $\Delta$$\lambda$, is approximately 1500. In each galaxy, between one and nine different spectral lines were observed over the 2.4-45 $\mu$m wavelength interval.

The data were reduced with the SWS Interactive Analysis package (Lahius et al. 1998, Wieprecht et al. 1998)  using the most recent set of calibration parameters and the {\it ISO} Spectral Analysis Package ISAP (Sturm et al. 1998).  Dark current subtraction, removal of cosmic ray glitches and noisy detectors, flat-fielding, and separate co-addition of data in each scan direction, was carried out prior to line profile extraction, and flux and uncertainty measurements.  In Table 2, we list line fluxes and upper limits for some of the most prominent lines within the SWS wavelength range. In some cases, observations for some of the galaxies exist at wavelengths that do not overlap with observations of any of the other galaxies; these line flux measurements were not included in Table 2.  Sample spectra of the high excitation [OIV] 25.6 $\mu$m spectral line, detected in 11 out of 18 LINERs observed at this wavelength, are shown for all previously unpublished observations in Figure 2.

\begin{figure}[h]
\noindent{\includegraphics[width=9cm]{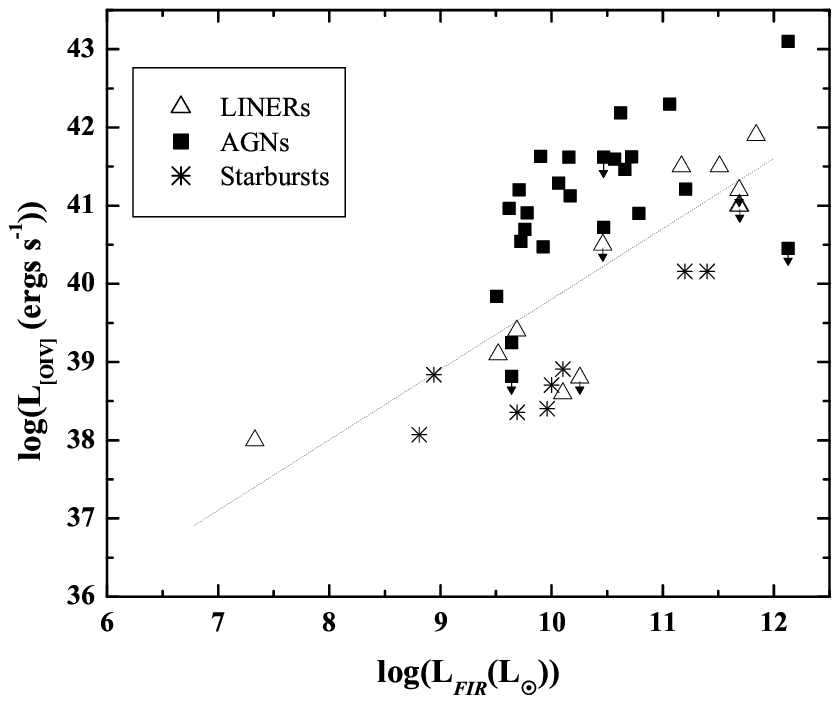}}
\caption[]{ The strength of the [OIV] emission line luminosity versus the total far-IR luminosity for LINERs compared with previously published AGN (Table 4) and starburst (Table 5) galaxies.  The dashed line corresponds to a linear fit to the LINER data.  As can be seen, LINERs fall between AGN and starbursts.}
\end{figure}

\subsection{\it Chandra Observations and Results}

Archival {\it Chandra} observations of all LINERs presented in this paper were obtained with the Advanced CCD Imaging Spectrometer (ACIS-S) with the source at the nominal aim point of the S3 CCD.  Most observations were carried out using the standard 3.2s  frame time, although a few bright targets were observed with shorter frame times.  Since many of the targets are nearby with bright nuclear X-ray emission, 10 out of the 33 LINER observations revealed pileup of the central unresolved source.  The measured counts per frame rate for the piled-up sources ranged from 0.1 to 0.97, making detailed spectral analyses unfeasible. 
 
	The {\it Chandra} observations were processed using CIAO v.2.1.2 using the latest calibration files provided by the {\it Chandra} X-ray Center (CXC).  In Table 3 we list the details of the {\it Chandra} observations along with our results; sources affected by pileup are flagged. Our data reduction procedure and analysis follow the treatment described by Ho et al. (2001).  The 0.2-8keV energy range was chosen for science analysis.  Nuclear count rates from 2-8 keV were extracted from a 2" radius centered on the nucleus.  For all galaxies that did not display dominant hard X-ray nuclear sources, the 2\,'' aperture was centered on the galaxy's {\it VLA} nucleus, if available, or the {\it 2MASS} or optical nucleus.  For most sources, the detected counts were insufficient to employ detailed spectral fits.    Therefore, the nuclear count rate was converted to 2-10 keV X-ray luminosities assuming an intrinsic power law spectrum with photon index $\Gamma$=1.8 using the Galactic interstellar absorption listed in Table 1.  For all observations that have been published as of 2003 January, the published intrinsic X-ray luminosities were always employed.  In most cases, these luminosities were taken from Ho et al. 2001, who employed the same method for obtaining intrinsic luminosities as we did.  Most of the sources that were effected by pile-up, were also previously published by authors who corrected for its effects.  In a few cases (Mrk 273, NGC 6240, NGC 5194), intrinsic absorption is known to significantly affect the intrinsic X-ray luminosities.  In these cases, published intrinsic X-ray luminosities obtained from detailed XSPEC fitting correcting for intrinsic absorption were employed.  Nuclear luminosities and references for published observations are listed in Table 3.

The X-ray images were classified using the morphological class designations adopted by Ho et al.  In their scheme, class (I) objects exhibit a dominant nuclear point source, class (II) objects exhibit multiple off-nuclear point sources of comparable brightness to the nuclear source, class (III) objects reveal a nuclear point source embedded in diffuse emission, and class (IV) objects display no nuclear source.  Morphological class designations for all galaxies are listed in Table 3.


\section{DISCUSSION OF RESULTS}

\subsection{\it Mid-IR Results: How do LINERs Compare with Starbursts and AGN?}

The mid-IR lines observed in our sample of LINERs originate from ions with a wide range of ionization potentials.  As has been shown in previous work, AGN show prominent high excitation fine structure line emission but starburst galaxies are characterized by a lower excitation spectra characteristic of HII regions ionized by young stars (e.g., Genzel et al. 1998, Sturm et al. 2002).  Of the 28 LINER galaxies, 18 were observed at the wavelength of the [OIV] 26 $\mu$m (excitation potential = 55 eV) line.  This line is not produced in HII regions surrounding young stars, the dominant energy source in starburst galaxies, since even hot massive stars emit very few photons with energy sufficient for the production of [OIV].  In Figure 3, we compare the strength of the [OIV] emission line relative to the total far-IR luminosity in LINERs, standard optically- identified AGN (Sturm et al. 2002) and the few starburst galaxies that show measurable [OIV] 26 $\mu$m fluxes (Lutz et al. 1998, Genzel et al. 1998).  The far-IR luminosity, a significant fraction of the bolometric luminosity of all of the galaxies plotted in Figure 3, comes from thermally reprocessed radiation from both stars and any possible AGN.  Since strong [OIV] emission requires the presence of an AGN, galaxies dominated by starbursts should  display low L$_{[OIV]}$/L$_{FIR}$ values (e.g., Genzel et al. 1998, Sturm et al. 2002).   A compilation of all previously published SWS data of AGN and starburst galaxies (plotted in Figure 3) is given in Tables 4 and 5.  Note that most starbursts do not show measurable [OIV] fluxes.  The origin of the line emission in the few starbursts plotted in Figure 3 is thought to be extended ionizing shocks related to the starburst activity  (Lutz et al. 1998), or, in the case of NGC 5253 and II Zw 40, the two starbursts with extreme ratios, photoionization by Wolf-Rayet stars (Schaerer \& Stasinska 1999).  As can be seen in Figure 3, LINERs show L$_{[OIV]}$/L$_{FIR}$ ratios intermediate between those of the AGN and starburst galaxies.  The average L$_{[OIV]}$/L$_{FIR}$ ratios are 4.2$\times$10$^{-5}$, 3.7$\times$10$^{-4}$, 3.4$\times$10$^{-3}$ for starbursts, LINERs, and AGN, respectively.


\begin{table*}

\begin{center}
\begin{tabular}{ccccccc}
\multicolumn{7}{l}{{\bf Table 4: AGN Comparative Sample}} \\
& &\\ \hline \hline
& & & & & &  \\
\multicolumn{1}{c}{Galaxy Name} & \multicolumn{1}{c}{Distance} & \multicolumn{1}{c}{L$_{FIR}$(L$_{\odot}$)} & F$_{[NeII]12.8{\mu}m}$$^a$  & F$_{[OIV]26{\mu}m}$$^a$ &  F$_{[SiII]35{\mu}m}$$^a$ & log(L$_{X}$) (2-10 keV) \\ 

\multicolumn{1}{c}{} & \multicolumn {1}{c}{Mpc} & \multicolumn{1}{c}{} & \multicolumn{1}{c}{10$^{-20}$ Wcm$^{-2}$} & \multicolumn{1}{c}{10$^{-20}$ Wcm$^{-2}$} & 10$^{-20}$ Wcm$^{-2}$ & ergs s$^{-1}$  \\ 
 
\multicolumn{1}{c}{(1)} & \multicolumn{1}{c}{(2)} & \multicolumn{1}{c}{(3)} & (4) & (5) & (6) & (7) \\
& & & & & &  \\ \hline
& & & & & & \\
Cen A & 4 & 4.36E+09 & 22.1 & 9.8 & 54.5 & $\cdots$ \\	
Circinus & 3 & 3.19E+09 & 90 & 67.9 & 151 & $\cdots$ \\
M 87 & 17 & 2.18E+08 & $\cdots$ & $<$0.2 & $<$1.2 & $\cdots$ \\	
Mkn 1 & 65 & 1.47E+10 & $\cdots$ & 2.8 & $<$1.3 & $\cdots$ \\
Mkn 3 & 54 & 1.43E+10 & 4.7 & 12.6 & $\cdots$ & 42.96$^b$ \\
Mkn 6 & 76 & $\cdots$ & 0.8 & 1.63 & $\cdots$ & $\cdots$ \\
Mkn 335 & 104 & 5.13E+09 & $<$0.7 & 1.3 & $\cdots$ & 43.4$^b$ \\
Mkn 463 & 202 & 1.15E+11 & 1.3 & 4.3 & 2 & 42.5$^c$ \\
Mkn 509 & 139 & 3.67E+10 & 2 & 1.8 & $\cdots$ & 44.39$^d$ \\
Mkn 573 & 69 & 7.96E+09 & $<$1.3 & 7.9 & $\cdots$ & $^b$ \\	
Mkn 938 & 80 & 1.43E+11 & 5.7 & $\cdots$ & $\cdots$ & $\cdots$ \\
Mkn 1014 & 677 & 1.34E+12 & 0.43 & 2.43 & $<$1.3 & $\cdots$ \\
NGC 613 & 20 & 1.59E+10 & 4 & $\cdots$ & $\cdots$ & $\cdots$ \\	
NGC 1068 & 14 & 5.24E+10 & 70 & 190 & 91 & 41.61$^b$ \\
NGC 1275 & 71 & 4.73E+10 & 2.9 & $<$0.5 & 6.8 & 43.81$^e$ \\
NGC 1365 & 22 & 6.09E+10 & 40.9 & 14.6 & 73.8 & 40.60$^f$ \\
NGC 3783 & 34 & 5.77E+09 & $\cdots$ & 3.8 & 3.2 & 43.32$^b$ \\
NGC 4051 & 10 & 1.57E+09 & 2.5 & $\cdots$ & $\cdots$ & 41.61$^b$ \\
NGC 4151 & 20 & 4.17E+09 & 11.8 & 20.3 & 15.6 & 43.07$^b$ \\
NGC 5506 & 23 & 6.02E+09 & 5.9 & 13astro-ph/0306510.5 & 14.2 & 43.31$^b$ \\
NGC 5643 & 16 & 8.43E+09 & $\cdots$ & 10.3 & 5.5 &$^b$ \\	
NGC 7469 & 65 & 1.60E+11 & 22.6 & 3.4 & 19.6 & 43.68$^b$ \\
NGC 7582 & 20 & 2.93E+10 & 14.8 & 11.6 & 21.8 & 41.9$^c$ \\
PKS 2048-57 & 45 & 1.15E+10 & 2.13 & 8.5 & 3.5 & $\cdots$ \\
TOL 0109-383 & 47 & 5.30E+09 & $<$1.2 & 1.4 & $\cdots$ & $\cdots$ \\	
I Zw 1 & 248 & 1.89E+11 & 0.65 & $<$0.6 & $<$1.8 & 43.39$^g$ \\
I Zw 92 & 153 & 4.58E+10 & 2.4 & 1.1 & $\cdots$ & $\cdots$ \\	
3C 120 & 135 & 4.17E+10 & 0.9 & 7.5 & $\cdots$ & 44.33$^b$ \\

& & & & & &  \\ \hline 
\end{tabular}
\end{center}
{\scriptsize{\bf Columns Explanation:} Col(1): Common Source Names ; Col(2): Distance in Mpc; Col(3):  Far-infrared luminosities correspond to the 40-500$\mu$m wavelength interval and were calculated using the IRAS 60 and 100 $\mu$m fluxes according to the prescription: L$_{[FIR]}$=1.26$\times$10$^{-14}$(2.58f$_{60}$+f$_{100}$) in W m$^{-2}$ (Sanders$\&$Mirabel 1996); Col(4): Flux of 12.8$\mu$m NeII line; Col(5):Flux of 26$\mu$m OIV line; Col(6):  Flux of 35$\mu$m SiII line; Col(7): Log of the X-ray luminosity in the 2-10 keV band}

{\noindent{\scriptsize{\bf References:} $^a$ Fluxes from Sturm et al. 2002; $^b$ Mas-Hesse et al. 1995; $^c$ Levenson et al. 2001; $^d$ Reynolds 1997; $^e$ Bassani et al. 1999; $^f$ Terashima et al. 2002 ApJS 139; $^g$ Leighly 1999}}
 
\end{table*}


\begin{table*}
\begin{center}
\begin{tabular}{cccccc}
\multicolumn{6}{l}{{\bf Table 5: Starburst Comparative Sample}} \\
& &\\ \hline \hline
& & & & &   \\
\multicolumn{1}{c}{Galaxy Name} & \multicolumn{1}{c}{Distance} & \multicolumn{1}{c}{L$_{FIR}$(L$_{\odot}$)} & F$_{[NeII]12.8{\mu}m}$$^a$  & F$_{[OIV]26{\mu}m}$$^b$  & log(L$_{X}$) (2-10 keV) \\ 

\multicolumn{1}{c}{} & \multicolumn {1}{c}{Mpc} & \multicolumn{1}{c}{} & \multicolumn{1}{c}{10$^{-20}$ Wcm$^{-2}$} & \multicolumn{1}{c}{10$^{-20}$ Wcm$^{-2}$} & ergs s$^{-1}$  \\ 

\multicolumn{1}{c}{(1)} & \multicolumn{1}{c}{(2)} & \multicolumn{1}{c}{(3)} & (4) & (5) & (6) \\
& & & & &   \\ \hline
& & & & & \\
N3256 & 37 & 1.75E+11 & 7.6 & 0.93 & 41.03$^d$ \\
M82 & 3 & 1.39E+10 & 88 & 8 & 39.53$^e$ \\
N3690A & 40 & 2.27E+11 & 3.2 & $<$1.2 & $\cdots$ \\	
N3690B & 40 & 2.27E+11 & 2.8 & 0.8 & $\cdots$ \\
N4038/39 & 21 & 2.91E+10 & 0.77 & $<$0.9 & $\cdots$ \\
N4945 & 4 & 9.21E+09 & 8.8 & 1.4 & $\cdots$ \\
M83 & 5 & 4.84E+09 & 13.4 & 0.8	& $\cdots$ \\
NGC 5253 & 4 & 6.51E+08 & 0.77 & 0.65 & $\cdots$ \\	
Galactic center & 0.008 & 1.50E+07 & $\cdots$ & $\cdots$ & $\cdots$ \\
N7469 & 66 & 1.65E+11 & 2.3 & $\cdots$ &  $\cdots$\\		
N7552 & 21 & 4.72E+10 & 6.8 & $<$1.2 &  $\cdots$\\
2 Zw 40 & 10.52 & 8.73E+08 & $\cdots$ & 0.55 & $\cdots$ \\	
IC 342 & $\cdots$ & $\cdots$ & $\cdots$& $<$1 & $\cdots$ \\	
NGC 6052 & 62.88 & 3.95E+10 & $\cdots$ & $<$0.8 & $\cdots$ \\
NGC 6764 & 32.21 & 1.08E+10 & $\cdots$ & $<$0.8 & $\cdots$ \\	

& & & & & \\ \hline 
\end{tabular}
\end{center}
{\scriptsize{\bf Columns Explanation:} Col(1): Common Source Names ; Col(2): Distance in Mpc; Col(3):  Far-infrared luminosities correspond to the 40-500$\mu$m wavelength interval and were calculated using the IRAS 60 and 100 $\mu$m fluxes according to the prescription: L$_{[FIR]}$=1.26$\times$10$^{-14}$(2.58f$_{60}$+f$_{100}$) in W m$^{-2}$ (Sanders$\&$Mirabel 1996); Col(4): Flux of 12.8$\mu$m NeII line; Col(5):Flux of 26$\mu$m OIV line; Col(6): Log of the X-ray luminosity in the 2-10 keV band}

{\noindent{\scriptsize{\bf References:} $^a$ Fluxes from Genzel et al. 2002; $^b$ Lutz 1998; $^c$Ptak et al. 1997; $^d$ Moran, Lehnert, \& Helfand 1999; $^e$ Ptak \& Griffiths 1999 and Matt et al. 1999.}}

\end{table*}


Figure 4 shows the L$_{[OIV]}$/L$_{FIR}$ ratio as a function of the IRAS F25$\mu$m/ F60$\mu$m flux ratio.  A warm mid-IR color indicates a hotter dust temperature and is a well-known powerful discriminator between AGNs and starforming galaxies (e.g. Miley, Neugebauer, \& Soifer 1985; de Grijp, Lub, \& Miley 1987, de Grijp et al. 1992).  It is clear from Figure 4 that while LINERs generally have higher L$_{[OIV]}$/L$_{FIR}$ ratios compared with starbursts, they have comparable mid-IR colors.  None of the LINERs presented here show extreme warm mid-IR colors characteristic of many AGN.
\begin{figure}[]
\noindent{\includegraphics[width=9cm]{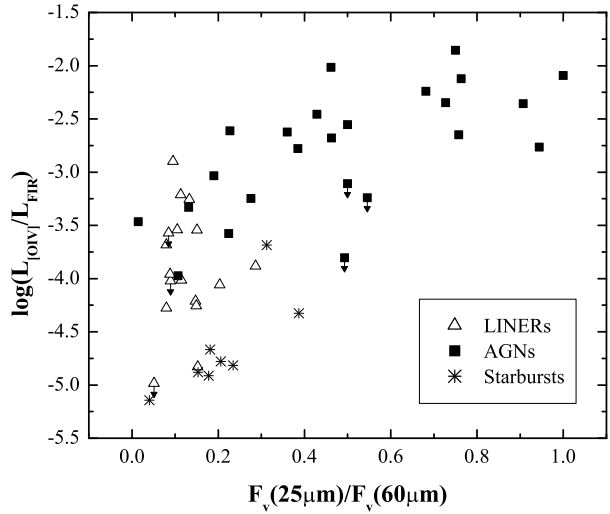}}\\
\caption[]{The [OIV] emission line to total far-IR luminosity ratio versus mid-IR IRAS color for LINERs compared with AGN and starburst galaxies.}
\end{figure}
\begin{figure}[h]
\noindent{\includegraphics[width=9cm]{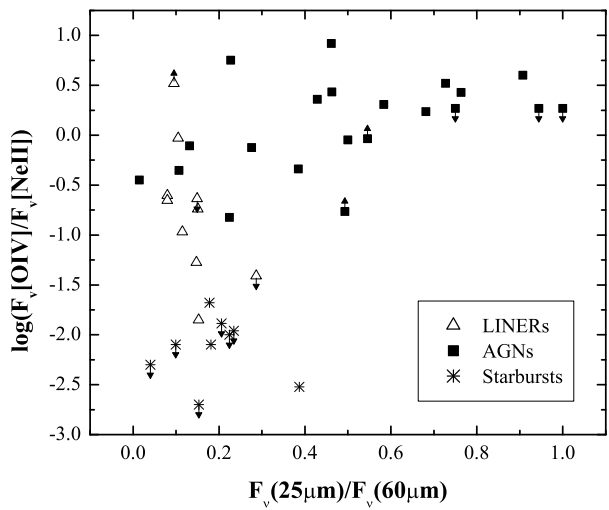}}\\
\caption[]{The [OIV] 26 $\mu$m / [NeII] 12.8 $\mu$m line flux ratio versus mid-IR IRAS color for LINERs compared with AGN and starburst galaxies.  Note that LINERs are intermediate between starbursts and AGN in their [OIV] / [NeII] line flux ratio.}
\end{figure}



In Figure 5, we plot the ratio of the high to low excitation line flux ratio [OIV] 26 $\mu$m/[NeII] 12.8 $\mu$m as a function of the IRAS 25/60 $\mu$m color for the 9 LINER galaxies for which observations at the wavelengths of both lines exist. As was first shown by Genzel et al. (1998), AGN show much higher [OIV]/[NeII] line flux ratios than starbursts.   The ionization potential of [NeII] is approximately 22 eV compared with 55 eV for [OIV].  Unlike the [OIV] emission, [NeII] emission arises in gas photoionized by both stars and AGN.  [NeII] emission in AGN narrow line regions is expected to be weak since the Neon is in higher ionization states.   In the absence of abundance variations and extreme extinction significant at mid-IR wavelengths, the [OIV]/[NeII] line flux ratio can be used as a measure of the dominant energy source in the galaxies presented here.  We can see from Figure 5 that LINERs show [OIV]/[NeII] ratios (average$=$0.28) intermediate between those of AGN (average$=$2.17) and starburst galaxies (average$=$0.02).  In addition, LINERs show the greatest dispersion in this ratio, possibly supporting the view that LINERs are a mixed bag of objects.  

Unfortunately, the sparse spectral emission line data for the LINERs observed by {\it ISO} precludes the ability to conduct detailed photoionization modeling of the central excitation source.
\begin{figure}[h]
\noindent{\includegraphics[width=9cm]{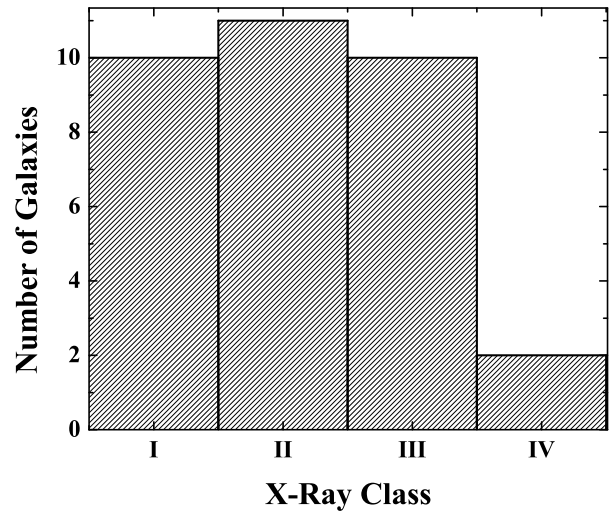}}
\caption[]{Demographics of the X-ray morphology for the sample of LINERs. Following the classification scheme from Ho et al. 2001, class I=dominant nucleus, class II=scattered sources, class III= nucleus embedded in soft diffuse emission, class IV=no nucleus (see text section 3.2 for details.}
\end{figure}

\begin{figure}[h]
\noindent{\includegraphics[width=9cm]{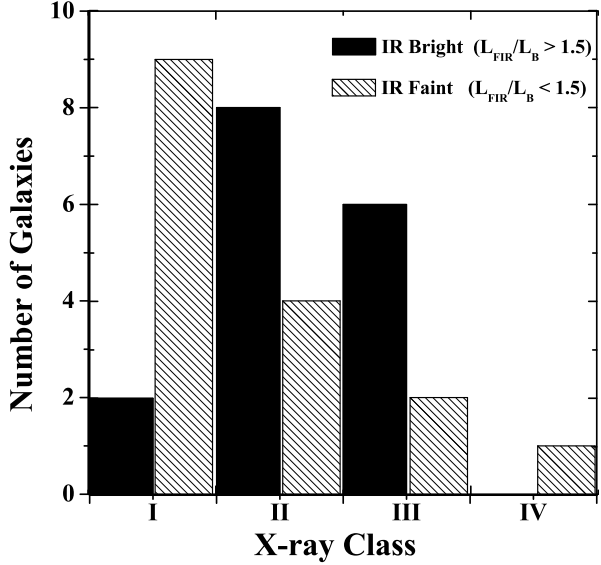}}
\caption[]{Demographics of the X-ray morphology for the sample of LINERs as a function of IR-brightness. (Classification scheme as in Figure 6.  See text for details).}
\end{figure}

\subsection{\it X-ray Morphology}

X-Ray observations can provide a powerful probe into the nuclei of obscured AGN out to column densities of a few times 10$^{24}$ cm$^{-2}$. While young supernova remnants, X-ray binaries, and/or hot diffuse gas from starburst-driven superwinds can contribute to the hard X-ray band, these sources of emission are generally weak and spatially extended.  Using {\it Chandra}'s high angular resolution (0.5"/pixel), detection of a compact hard X-ray source in these nearby LINERs is highly suggestive of an AGN.  Note that for 70\% of the sample, 0.5" corresponds to less than 100 pc.  Figure 6 shows the distribution of X-ray morphology amongst the sample of LINERS.  As can be seen, we detect nuclear point sources morphologically consistent with AGN in approximately 67\% of the sample. 


\begin{figure}[]
\noindent{\includegraphics[width=9cm]{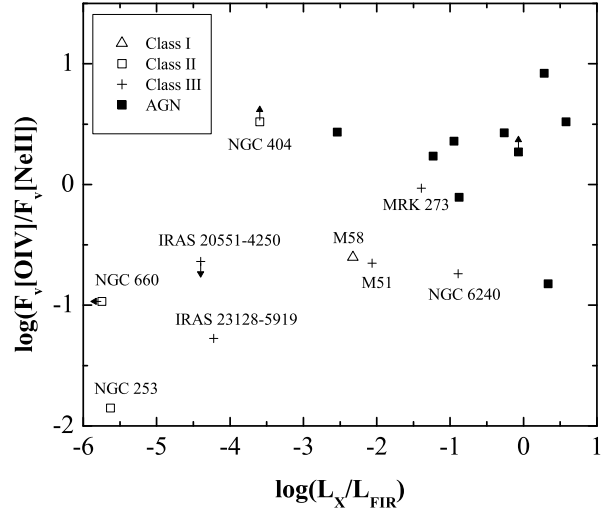}}
\caption[]{The [OIV]26 $\mu$m / [NeII]12.8 $\mu$m line flux ratio versus hard (2-10 keV) X-ray to far-IR luminosity ratio for LINERs compared with AGN.  LINERs are labeled by galaxy name with symbols denoting X-ray morphological class designations.  (see text section 3.2 for details).}
\end{figure}
\begin{figure}[]
\noindent{\includegraphics[width=9cm]{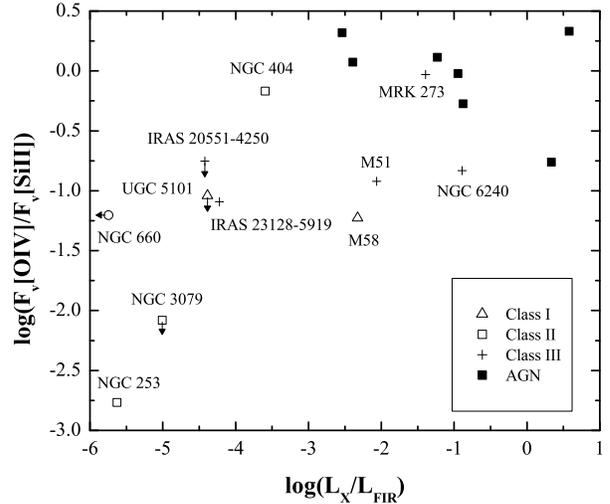}}
\caption[]{The [OIV]26 $\mu$m / [SiII]34 $\mu$m line flux ratio versus hard (2-10 keV) X-ray to far-IR luminosity ratio for LINERs compared with AGN.  Symbols as in Figure 8.}
\end{figure}



In order to explore the characteristics of the IR-bright LINER population, we compared the X-ray morphologies of LINERs with L$_{FIR}$/L$_{B}$ $>$ 1 with those with L$_{FIR}$/L$_{B}$ $<$ 1.  As can be seen in Figure 7, most (82\%) LINERs that show a single dominant compact X-ray core (class I) are IR-faint.  On the other hand, most (64\%) LINERs that show scattered X-ray sources (class II) are IR-bright.  Several LINERs show hard compact X-ray cores embedded in diffuse soft X-ray emission (class III); virtually all of these LINERs are IR-bright.  

These trends of morphological classification with IR-brightness are consistent with expectations.  Class II objects are morphologically consistent with starburst galaxies, expected to contain a large population of young stars that heat the dust, as well as a large population of X-ray binaries consistent with an X-ray morphology dominated by scattered discrete sources.  Class III objects are very likely objects that contain both dust enshrouded circumnuclear starformation as well as an embedded AGN. Indeed, obscuration of AGN is invariably accompanied by starburst activity in Seyfert 2 galaxies (Maiolino et al. 1995).  


\subsection{\it Mid-IR/X-ray Comparative Study: Multiwavelength Consistency?}

Do multiwavelength diagnostics result in a self-consistent understanding of LINERs?  Objects that display strong hard X-ray nuclear point sources should also display high excitation lines in the IR.  However, for the few IR-bright LINERs studied in detail at both X-ray and IR wavelengths prior to this work, a conflicting picture emerges.  The IR-bright LINERs NGC 6240 and Mrk 273 both show weak [OIV] 26 $\mu$m emission relative to the emission from lower excitation forbidden lines (Genzel et al. 1998), but both show remarkably luminous hard X-ray nuclear point sources (e.g., Komossa \& Shultz 1999, Xia et al. 2002).  If these LINERs are indeed powered by AGN, why do they not display strong high excitation lines at mid-IR wavelengths where dust obscuration is minimal? Do other LINERs show similar multiwavlength inconsistencies?

Figure 8 shows the  [OIV]/[NeII] line flux ratio as a function of the hard X-ray to far-IR luminosity ratio for this sample of LINERs compared with those AGN for which both SWS and X-ray observations exist (see Table 5 for references). Like the [OIV] emission, the hard X-ray emission is dominated by the AGN.   The first thing to note from Figure 8 is that, as expected, the galaxy with the lowest  [OIV]/[NeII] line flux ratio is X-ray weak and morphologically consistent with a starburst in the X-rays (class II object).  Galaxies with the highest L$_{X}$/L$_{FIR}$ ratios, are all morphologically consistent with AGN in the X-rays (class I and III objects).  However, there are a few anomalies.  NGC 6240 is clearly extremely luminous and morphologically consistent with an AGN in he hard X-rays but, as pointed out by Genzel et al. (1998), the mid-IR emission line ratios do not suggest an energetically-dominant AGN.  We note that the Seyfert 1 galaxy NGC 7469 also shows strong hard X-ray emission from a nuclear point source but an anomalously low [OIV]/[NeII] line flux ratio.  Both galaxies are IR-bright and are known to host vigorous star formation activity (e.g. Genzel et al. 1995, van der Werf et al. 1993).  Extinction estimates are insufficient to affect substantially the mid-infrared line ratio suggesting that some other explanation for this apparent multiwavelength inconsistency is required.  If the intrinsic spectral energy distribution from the X-ray to UV wavelength band is consistent with a standard AGN in these two galaxies, the emission lines from highly ionized gas should dominate their emission line spectra.  In contrast, NGC 404 shows very weak soft X-ray emission consistent with a starburst (see Eracleous et al. 2002) but has the highest [OIV]/[NeII] ratio of our entire sample.  Using IR emission line diagnostics alone, this galaxy would be classified as hosting an energetically-dominant AGN.  If the emission lines are produced in photoionized plasma surrounding X-ray binaries, it is indeed surprising that the forbidden line luminosities from the high excitation ions surpass those produced by the AGN plotted in Figure 8.  In Figure 9, we plot a similar diagnostic diagram using the [OIV] 26 $\mu$m / [SiII] 34 $\mu$m line flux ratio.  Since the ionization potential of Silicon is less than that of Hyrdogen, [SiII] 34 $\mu$m emission can originate from both photoionized and photodissociated gas.  Due to the potentially higher importance of partially ionized zones in AGNs compared with starbursts, the  diagnostic power of this line ratio is more ambiguous.  Nonetheless,  it is clear that Figure 9 displays the same trends and anomalies that are seen in Figure 8.  We stress that our choice of IR line ratio diagrams is limited by the paucity of data available for LINERs.  With the advent of {\it SIRTF}, detailed abundance- and extinction-insensitive photoionization modeling of the excitation source in LINERs will for the first time be possible.

\subsection{\it {Transient Phenomena in LINERs?}}

The multiwavelength anomalies pointed out above invoke the possibility of transient activity in at least some LINERs. A low X-ray luminosity and a prominent high-excitation forbidden line spectrum could be produced if both phenomena are variable and the {\it ISO} and {\it Chandra} observations took place on different dates.  Similarly, a luminous hard X-ray luminosity morphologically consistent with an AGN could be reconciled with an anemic forbidden line spectrum if the AGN activity is intermittent and the observations were not concurrent.  Several LINERs show broad variable Balmer lines (e.g., Storchi-Bergmann, Baldwin, \& Wilson 1993, Bowler et al. 1996, Eracleous \& Halpern 2001), although variability in the narrow lines has not been seen.  X-ray variability on timescales of weeks to several years has also been observed (Terashima et al. 2002).  The {\it ISO} observations of NGC 404 were obtained during 1998 January, approximately two years prior to the {\it Chandra} obserservations. However, NGC 404 was detected in the soft X-rays  by ROSAT HRI in 1997 April and found to be weak (Komossa, Bohringer, \& Huchra 1999).   An upper limit by ASCA in 1997 July also indicates that it was X-ray weak in the hard band indicating that X-ray variability between the dates of the {\it ISO} and {\it Chandra} observations is unlikely.  Similarly, prior to the recent detection of the binary AGN in NGC 6240 by {\it Chandra} (Komossa et al. 2003), the presence of a highly absorbed extremely X-ray luminous AGN was indicated (e.g., Vignati et al. 1999).  In addition, the apparent deficit of emission lines from ionized gas implied by the {\it ISO} spectrum was also indicated by the low hydrogen recombination line fluxes observed from NGC 6240 more than a decade ago (e.g., DePoy, Becklin, \& Wynn-Williams 1986; Thronson et al. 1990; van der Werf et al. 1993).  It is therefore highly unlikely that the multiwavelength anomalies in these galaxies can be explained by variability.

\subsection{\it { Constraints on the Intrinsic Spectral Energy Distribution in LINERs: A UV Deficit?}}

A luminous hard X-ray luminosity morphologically consistent with an AGN in concert with weak forbidden and recombination lines invites the possibility that the intrinsic spectral energy distribution (SED) in at least some LINERs is different than that of standard AGN. Previous observations of the continuum emission in several low luminosity LINER galaxies reveal that the observed SEDs from the nuclear source lacks the standard "big blue bump" in the UV, a feature normally associated with emission from a standard optically thick geometrically thin accretion disk (Lasota et al. 1996; Ho 1999).  At low accretion rates, the accretion is thought to take place via advection-dominated accretion flows (ADAFs; Narayan \& Yi 1994, 1995).  Since ADAFs are radiatively inefficient in the production of UV photons, this scenario was suggested by Ho (1999) as the explanation for the UV photon deficit in the LINER SEDs.

However, recent observations of a few galaxies that show LINER-like optical spectra suggest alternative scenarios.  For example, detailed photoionization modeling of the LINER-like optical emission lines in the weak line radio galaxy NGC 4261 require a substantially stronger ionizing UV continuum than is observed.  This suggests that the intrinsic SED in this galaxy is similar to standard AGN and the observed UV deficiency in the continuum is a result of substantial extinction along our line of sight to the nucleus.  This is supported by recent {\it XMM} and {\it Chandra} studies that reveal absorption of N$_{H}$ $\sim$ 10$^{22}$cm$^{-2}$ towards the nucleus (Sambruna et al. 2003; Gliozzi et al. 2003).  Furthermore, The variablity trends in NGC 4261 are very similar to the standard disk-corona system of Seyferts, which may suggest that the X-ray continuum is not likely to originate in an ADAF (Gliozzi et al. 2003).  In this section, we investigate the intrinsic SEDs of our sample of LINERs implied by our IR emission line observations.

  The infrared emission lines presented here are powerful probes of the intrinsic ionizing radiation field as seen by the narrow line region.  Unlike optical emission lines, these lines are less sensitive to extinction.  Figure 10 shows the [OIV] 26 $\mu$m luminosity versus the hard X-ray luminosity for all LINERs and standard AGN.  According to Ho (1999), the UV band is exceptionally dim relative to both the optical and X-ray bands for seven low luminosity LINERs including NGC 4579, one of the LINERs plotted in Figure 10.  However, as can be seen by Figure 10, the correlation between the [OIV] 26 $\mu$m and hard X-ray emission for virtually all LINERs and AGN strongly suggests a similar intrinsic SED between the UV and X-ray bands.  
	Figure 11 shows the [NeV] 14.3 $\mu$m / [OIV] 26 $\mu$m  line flux ratio versus the hard X-ray luminosity for LINERs and AGN.  The ionization potential of [NeV] is 97 eV.  Since both lines are produced predominantly by AGN, the line ratio can be used to probe the shape of the ionizing radiation field and ionization parameters characteristic of the narrow line region.  As can be seen from Figure 11, the line ratios characteristic of LINERs fall well within the range observed for AGN.  In fact, the line ratios do not require low ionization parameters, one of the main hypotheses for the low excitation optical line spectrum in LINERs (Ferland \& Netzer 1983; Halpern \& Steiner 1983).  

\begin{figure}[]
\noindent{\includegraphics[width=9cm]{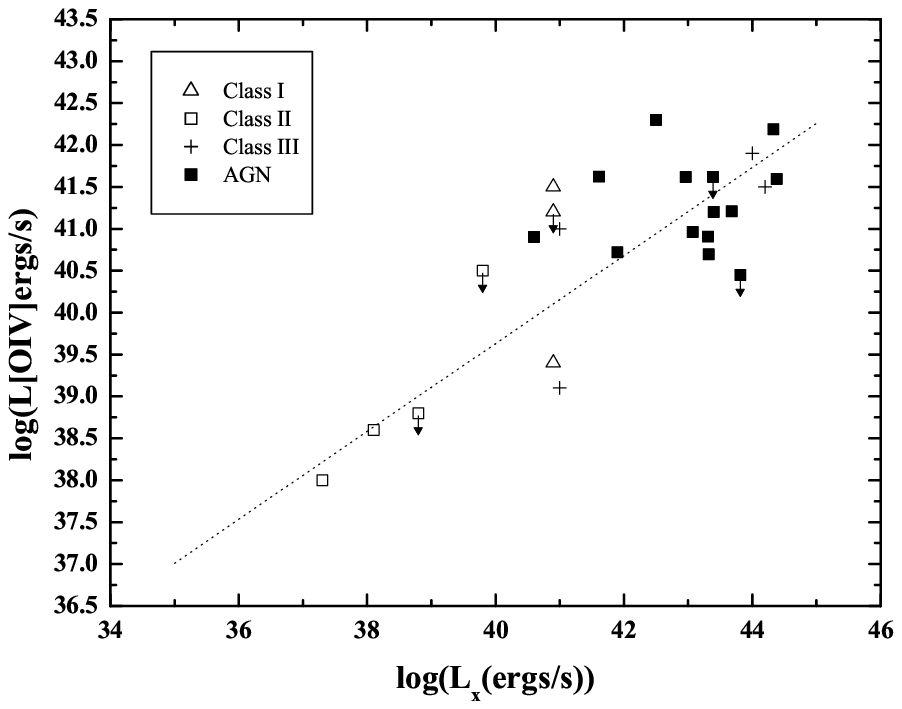}}
\caption[]{The [OIV] 26 $\mu$m luminosity versus the hard (2-10 keV) X-ray luminosity for LINERs compared with AGN.  The dotted line corresponds to a linear fit to the LINER data.  Symbols as in Figure 8.}
\end{figure}


\begin{figure}[]
\noindent{\includegraphics[width=9cm]{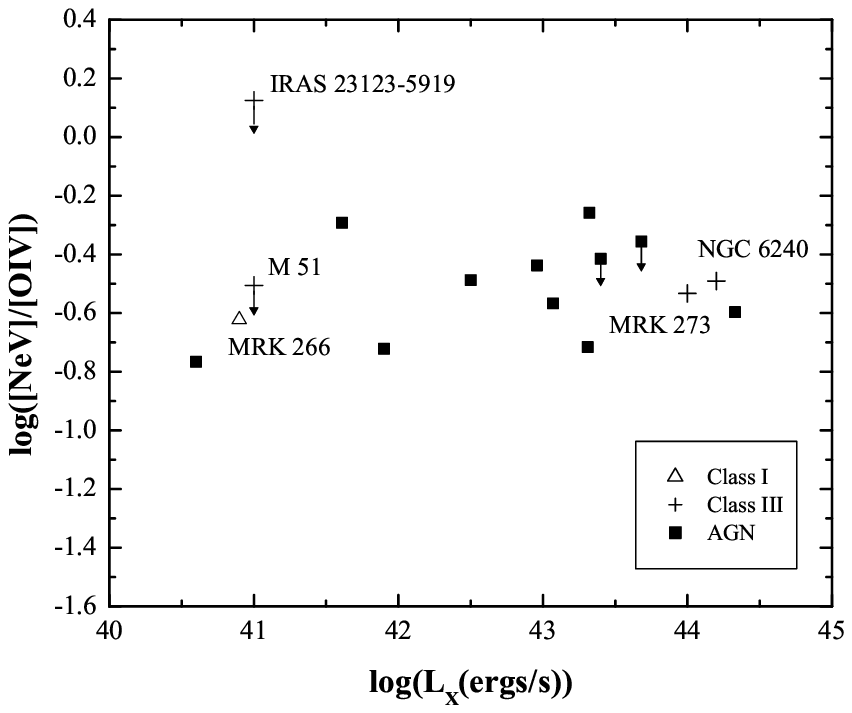}}
\caption[]{The [OIV] 26 $\mu$m luminosity versus the hard (2-10 keV) X-ray luminosity for LINERs compared with AGN.  The dotted line corresponds to a linear fit to the LINER data.  Symbols as in Figure 8.}
\end{figure}


In summary, although the data is sparse, our preliminary IR observations suggest that the intrinsic SEDs in LINERs are not necessarily different from those of standard AGN.  While NGC 6240, possesses an emission line spectrum in the IR that may suggest a deficit of UV photons, its hard X-ray luminosity is comparable to some of the most luminous AGN and is unlikely to be fueled by the low mass accretion rates thought to be associated with ADAFs.  Furthermore, several bona fide Seyfert 1 galaxies also display similar low excitation IR spectra suggesting a common origin for this phenomena.  Since the two X-ray luminous galaxies with weak emission lines are both IR luminous, it is very likely that dusty AGN are responsible for their observed characteristics.   For example, it is possible that dust absorption of ionizing radiation and subsequent direct conversion to thermal IR radiation is responsible for the weak emission line spectrum from the narrow line regions of some dust-enshrouded AGN (Netzer \& Laor 1993).  We stress that for a true understanding of the intrinsic SEDs of dusty LINERs, an extensive set of emission lines are required for detailed photoionization models.

\section{SUMMARY AND CONCLUSIONS}
We present a comprehensive comparative high resolution mid-IR spectroscopic and X-ray imaging investigation of all LINERs observed by {\it ISO} and {\it Chandra}.  Although the sample is heterogenous and incomplete, this is the first comprehensive study of the mid-infrared fine structure line emission of LINERs.  These results have been compared with similar observations of starburst galaxies and AGN. Our main results are summarized below:

1.	The high excitation [OIV] 26 $\mu$m spectral line, generally detected only in active galactic nuclei, was observed in 17 of the LINERs and detected in 11 of these objects.

2.	LINERs very clearly fall between starbursts and AGN in their mid-IR fine structure line spectra.  They show L$_{[OIV]26{\mu}m}$/L$_{FIR}$ and L$_{[OIV]26{\mu}m}$ / L$_{[NeII]12.8{\mu}m}$ ratios intermediate between those of AGN and starburst galaxies.  In the absence of abundance variations, these ratios are a measure of the dominant energy source in dusty galaxies.  The average L$_{[OIV]26{\mu}m}$ / L$_{FIR}$ ratio is 4.2 $\times$ 10$^{-5}$, 3.7 $\times$ 10$^{-4}$, and 3.4 $\times$ 10$^{-3}$ for starbursts, LINERs, and AGN, respectively. The average L$_{[OIV]26{\mu}m}$ / L$_{[NeII]12.8{\mu}m}$ ratio is 0.02, 0.28, and 2.17 for starbursts, LINERs, and AGN, respectively. In addition, LINERs show the greatest dispersion in these ratios, possibly supporting the view that LINERs are a mixed bag of objects.

3.	An X-ray morphological study of the 33 LINER galaxies observed by {\it Chandra} for which joint {\it ISO} comparison data is available has also been conducted.  We detect nuclear point sources morphologically consistent with AGN in ~ 67\% of the sample. We find that most (82\%) LINERs that show a single dominant hard compact X-ray core are IR-faint (L$_{FIR}$/L$_{B}$ $<$ 1).  On the other hand, most (64\%) LINERs that show scattered X-ray sources are IR-bright.  Several LINERs show hard compact X-ray cores embedded in diffuse soft X-ray emission; virtually all of these LINERs are infrared-bright. 

4.	A comparative X-ray/mid-IR spectroscopic investigation of LINERs reveals some surprising results.  Objects that display strong hard nuclear X-ray cores should also display high excitation lines in the IR.  However, the galaxy NGC 404 shows weak soft X-ray emission consistent with a starburst but has the highest excitation mid-IR spectrum of our entire sample.  Using IR emission line diagnostics, this galaxy would be classified as hosting a dominant AGN.  Conversely, the IR luminous LINER NGC 6240 has an extremely luminous binary AGN as revealed by the X-rays but shows weak IR emission lines.  

5.	Contrary to previous suggestions, a comparison of IR emissions lines with hard X-ray luminosities strongly suggest that the intrinsic SEDs in LINERs are not necessarily different in LINERs compared with standard AGN.\\

  Our joint mid-infrared spectroscopic and X-ray investigation of LINERs provide a sampling of some of the puzzles emerging concerning their central energy sources.  We stress that the limited data available at this time precludes detailed photoionization modeling of the central engines.  With the advent of {\it SIRTF}, and future IR missions such as Herschel and JSWT, such detailed analysis will for the first time be possible.\\

{\bf ACKNOWLEDGEMENTS}
We are very thankful to John McNulty for his assistance in analyzing the {\it Chandra} images.  We are also very grateful for the generous help from Jessica Gambill, Mario Gliozzi, Davide Donato, Hana Sanei, and the very helpful suggestions from the referree.  This research has made use of the NASA/IPAC Extragalactic Database (NED) which is operated by the Jet Propulsion Laboratory, California Institute of Technology, under contract with the National Aeronautics and Space Administration.  SS gratefully acknowledges financial support from NASA grant NAG5-11432.  RMS gratefully acknowledges financial support from NASA LTSA grant NAG5-10708 and from the Clare Boothe Luce Program of the Henry Luce Foundation.


\begin{thebibliography}{}

\bibitem[]{} Barger, A. J.; Cowie, L. L.; Mushotzky, R. F. \& Richards, E. A., 2001, AJ, 121, 662

\bibitem[]{} Bassani, L.; Dadina, M.; Maiolino, R.; Salvati, M.; Risaliti, G.; della Ceca, R.; Matt, G. \& Zamorani, G., 1999,  ApJS, 121, 473

\bibitem[]{} Bower, Gary A.; Wilson, Andrew S.; Heckman, Timothy M.; \& Richstone, Douglas O., 1996, AJ, 111, 1901

\bibitem[]{} Brandt et al., 2001, AJ, 122, 2810 

\bibitem[]{} Carrillo, R., Masegosa, J.,
 Dultzin-Hacyan, D., \& Ordonez, R. 1999, Revista Mexicana de Astronomia y Astrofisica, Vol. 35, p.187

\bibitem[]{} Comastri, A.; Mignoli, M.; Ciliegi, P.; Severgnini, P.; Maiolino, R.; Brusa, M.; Fiore, F.; Baldi, A.; Molendi, S.; Morganti, R.; \& 4 coauthors, 2002, ApJ, 571, 771

\bibitem[]{} de Graauw, T.; Haser, L. N.; Beintema, D. A.; Roelfsema, P. R.; van Agthoven, H.; Barl, L.; Bauer, O. H.; Bekenkamp, H. E. G.; Boonstra, A.-J.; Boxhoorn, D. R.; \& 52 coauthors, 1996, A \& A, 315, 49

\bibitem[]{} de Grijp, M. H. K.; Lub, J. \& Miley, G. K., 1987, A \& AS, 70, 95

\bibitem[]{} de Grijp, M. H. K.; Keel, W. C.; Miley, G. K.; Goudfrooij, P. \& Lub, J., 1992, A \& AS, 96, 389D

\bibitem[]{} Depoy, D. L.; Becklin, E. E. \& Wynn-Williams, C. G., 1986, ApJ, 307, 116

\bibitem[]{} Eracleous, Michael; Shields, Joseph C.; Chartas, George; \& Moran, Edward C.,  2002, ApJ, 565, 108

\bibitem[]{} Eracleous, Michael \& Halpern, Jules P., 2001, ApJ, 554, 240

\bibitem[]{} Feldmeir, John J. \& Ciardullo, Robin; 1997, ApJ, 479, 231

\bibitem[]{} Ferland, G. J.; \& Netzer, H., 1983, ApJ, 264, 105

\bibitem[]{} Fukazawa, Y.; Iyomoto, N.; Kubota, A.; Matsumoto, Y. \& Makishima, K., 2001, PASJ, 53, 595

\bibitem[]{} Filippenko, A. V., 1996, tpol.conf, 17

\bibitem[]{} Genzel, R.; Weitzel, L.; Tacconi-Garman, L. E.; Blietz, M.; Cameron, M.; Krabbe, A.; Lutz, D. \&  Sternberg, A., 1995, ApJ, 444, 129

\bibitem[]{} Genzel, R., et al. 1998, ApJ, 498, 579

\bibitem[]{} Gliozzi, M.; Sambruna, R. M.; \& Brandt, W.N., 2003, [astro-ph/0306510]

\bibitem[]{} Guainazzi, M.; Oosterbroek, T.; Antonelli, L. A. \& Matt, G. 2000, A \& A 364, 80

\bibitem[]{} Halpern, J. P.; \& Steiner, J. E., 1983, ApJ, 269, 37

\bibitem[]{} Heckman, T.M. 1980, A \& A, 87, 152

\bibitem[]{} Ho, L.C., Filippenko, A.V. \& Sargent, W.L.W. 1993, ApJ, 417, 63

\bibitem[]{} Ho, L.C., Filippenko, A.V. \& Sargent, W.L.W. 1997, ApJS, 112, 315 (HO97)

\bibitem[]{} Ho, L. C., Filippenko, A. V., Sargent, W. L. W., and Peng, C. Y., Astrophys. J. Suppl., 112, 391 (1997b)

\bibitem[]{} Ho, Luis C., 1999, AdSpR, 23, 813

\bibitem[]{} Ho, L. C., et al. 2001, ApJ 549, L51

\bibitem[]{} Kessler, M. F.; Steinz, J. A.; Anderegg, M. E.; Clavel, J.; Drechsel, G.; Estaria, P.; Faelker, J.; Riedinger, J. R.; Robson, A.; Taylor, B. G.; Ximenez de Ferran, S., 1996, A \& A, 315, 27

\bibitem[]{} Komossa, Stefanie \& Schulz, Hartmut, 1999, Ap \& SS, 266, 61

\bibitem[]{} Komossa, Stefanie; B\"{o}hringer, Hans \& Huchra, John P., 1999, A \& A, 349, 88

\bibitem[]{}   Komossa, Stefanie;  Burwitz, Vadim; Hasinger, Guenther; Predehl, Peter;  Kaastra,Jelle S.; \& Ikebe, Yasushi, 2002, [astro-ph/0212099]

\bibitem[]{} Komossa, S.; Burwitz, V.; Hasinger, G.; Predehl, P.; Kaastra, J. S. \& Ikebe, Y., 2003, ApJ, 582, 15

\bibitem[]{} Keil R., Boller, T., \& Fujmoto, R., 2001, [astro-ph/0106195] 

\bibitem[]{} Lahuis, F.; Wieprecht, E.; Bauer, O. H.; Boxhoorn, D.; Huygen, R.; Kester, D.; Leech, K. J.; Roelfsema, P. R.; Sturm, E.; Sym, N. J.; \& Vandenbussche, B., 1998, adass, 7, 224

\bibitem[]{} Lasota, J.-P.; Abramowicz, M. A.; Chen, X.; Krolik, J.; Narayan, R.; \&  Yi, I., 1996, ApJ, 462, 142

\bibitem[]{} Leighly, K. M., 1999, ApJS, 125, 317 

\bibitem[]{} Levenson, N. A.; Weaver, K. A.; \& Heckman, T. M., 2001, ApJS, 133, 269

\bibitem[]{} Lutz, D.; Spoon, H. W. W.; Rigopoulou, D.; Moorwood, A. F. M.; \& Genzel, R.,  1998, ApJ, 505, 103

\bibitem[]{} Lutz D., Veilleux. S., \& Genzel, R. 1999, ApJ, 517, L13 

\bibitem[]{} Maiolino, R.; Ruiz, M.; Rieke, G. H.; \& Keller, L. D., 1995, ApJ, 446, 561

\bibitem[]{} Moran, Edward C.; Lehnert, Matthew D. \& Helfand, David J., 1999, ApJ, 526, 629M

\bibitem[]{} Mas-Hesse, J. M.; Rodriguez-Pascual, P. M.; Sanz Fernandez de Cordoba, L.; Mirabel, I. F.; Wamsteker, W.; Makino, F. \& Otani, C., 1995, A \& A 298, 22

\bibitem[]{} Miley, G. K.; Neugebauer, G. \& Soifer, B. T., 1985, ApJ, 293, 11

\bibitem[]{} Mushotzky, R.F., Cowie, L.L., Barger, A.J., Arnaud, K.A., 2000, Nat, 404, 459

\bibitem[]{} Narayan, Ramesh; \& Yi, Insu, 1994, ApJ, 428L, 13

\bibitem[]{} Narayan, Ramesh; \& Yi, Insu, 1995, ApJ, 452, 710

\bibitem[]{} Netzer, Hagai; \& Laor, Ari, 1993, ApJ, 404, 51

\bibitem[]{} Prieto, M. Almudena \& Viegas, Sueli M., 2000, ApJ, 532, 238

\bibitem[]{} Puche, D.; Zijlstra, A. A.; Boettcher, C.; Plante, R. L.; Wilcots, E. M.; Wilkin, F. P.; Krause, S.; Sergo, S. P.; Bierman, G. S. \& Ge, J. 1988, 206, 89

\bibitem[]{} Ptak, A. \& Griffiths, R., 1999, ApJ, 517, 85

\bibitem[]{} Ptak, A.; Serlemitsos, P.; Yaqoob, T.; \& Mushotzky, R., 1997, AJ, 133, 1286

\bibitem[]{} Reynolds, C.S., 1997 MNRAS, 286, 513

\bibitem[]{} Sambruna, R. M.; Gliozzi, M.; Eracleous, M.; Brandt, W. N.; Mushotzky, 2003, ApJ, 586, 37

\bibitem[]{} Sanders, D.B. \& Mirabel, I.F., 1996, ARA \& A, 34, 749

\bibitem[]{} Satyapal, S.; \& Sanei, S., 2003, in prep.

\bibitem[]{} Schaerer, Daniel \& Stasinska, Grazyna, 1999, A\&A, 345, 17

\bibitem[]{} Stauffer, J.R. 1982, ApJS, 50, 517

\bibitem[]{} Stanek, K. Z. \& Garnavich, P. M. 1998, ApJ, 503, 131

\bibitem[]{} Storchi-Bergmann, Thaisa; Baldwin, Jack A.; \& Wilson, Andrew S., 1993, ApJ, 410, 11

\bibitem[]{} Sturm, E.; Bauer, O. H.; Brauer, J.; Buckley, M.; Harwood, A.; Helou, G.; Khan, I.; Li, J.; Lord, S.; Lutz, D.; \& 10 coauthors, 1998, adass, 7, 161

\bibitem[]{} Sturm, E.; Lutz, D.; Verma, A.; Netzer, H.; Sternberg, A.; Moorwood, A. F. M.; Oliva, E. \&  Genzel, R., 2002, A \& A, 393, 821

\bibitem[]{} Sugai, H. \& Malkan, M. A., 2000, ApJ, 529, 219

\bibitem[]{} Terashima, Y.; Iyomoto, N.; Ho, Luis C.; \& Ptak, A. F., 2002, ApJS, 139, 1 

\bibitem[]{} Thornley, Michele D., Schreiber, Natascha M. F\"{o}rster, Lutz, Dieter, Genzel, Reinhard, Spoon, Henrik W. W., Kunze, Dietmar,\&  Sternberg, Amiel 2000, ApJ, 539, 641

\bibitem[]{} Thronson, Harley A., Jr.; Majewski, S.; Descartes, Lara; \& Hereld, Mark, 1990, ApJ, 364, 456

\bibitem[]{} Tully, R. B., 1988, AJ, 96, 73

\bibitem[]{} van der Werf, Paul P.; Genzel, R.; Krabbe, A.; Blietz, M.; Lutz, D.; Drapatz, S.; Ward, Martin J.; Forbes \&  Duncan A., 1993, ApJ, 405, 522

\bibitem[]{} Veilleux, S., Sanders, D.B. \& Kim, D.-C. 1999, ApJ, 522, 139

\bibitem[]{} Vignati, P., et al. 1999, A \& A, 349, L57

\bibitem[]{} Wieprecht, E.; Lahuis, F.; Bauer, O. H.; Boxhoorn, D.; Huygen, R.; Kester, D.; Leech, K. J.; Roelfsema, P.; Sturm, E.; Sym, N. J.; \& Vandenbussche, B., 1998, adass, 7, 279

\bibitem[]{} Wilson, A. S. \& Yang, Y., 2001, AAS, 19915005

\bibitem[]{} Xia, X. Y. \& Xue, S. J., 2001, tysc.confE, 40

\bibitem[]{} Xia, X. Y.; Xue, S. J.; Mao, S.; Boller, Th.; Deng, Z. G.; \& Wu, H., 2002, ApJ, 564, 196

\end{thebibliography}
\end{document}